\documentclass[11pt,letterpaper]{article}
\usepackage[margin=1in]{geometry}
\usepackage[T1]{fontenc}
\usepackage[utf8]{inputenc}
\usepackage{mathptmx,amsmath,amssymb,graphicx,booktabs,array,caption,float}
\usepackage[hidelinks]{hyperref}
\makeatletter\renewcommand\@biblabel[1]{}\makeatother
\title{AI and the Two Margins of Work:\\Why Job Losses Show Up Before Job Creation}
\author{Mark Esposito\\{\normalsize micro1 and Northeastern University} \and Liu Zhang\\{\normalsize micro1}}
\date{November 2026}
\begin{document}
\maketitle
\begin{center}\textit{}\end{center}
\begin{abstract}
Technological change acts on labor demand at two margins: adjustment within established activities and the creation of additional activities, often by new firms. Payroll statistics record the first promptly and the second late. In the computing era, firms that entered in 1978 or later held 42 percent of payroll by 1998 and entering cohorts kept about 80 percent of their payroll after five years; in the industries most exposed to automation, incumbents contracted before entrants accumulated, a sequence that is precise early in the period and suggestive by its end. We develop a model in which one AI capability lowers the full cost of acceptable output in existing production and makes additional projects worth attempting, with founder effort separate from hired hours. If operating products re-scale instantly, automation displaces no aggregate hours; if they take time, hours fall first and recover as incumbents re-expand and new products arrive, with the long-run sign set by income effects. Establishment job flows through the first quarter of 2026 show no gradient on language-model exposure on either margin, which is what the model predicts under the observed adoption path and also what no effect on payroll would look like; studies with founding-task exposure find entry rising 10 to 20 percent. We propose a net reinstatement ratio and show that statistics which record new work late overstate displacement for years.
\end{abstract}
\vspace{6pt}
\section{Introduction}

Every major technological revolution has arrived with the same forecast: that machines would make human labor redundant. The Luddites smashed stocking frames in 1811, Keynes (1930) warned of technological unemployment, and mid-century commissions in the United States debated automation as a threat to full employment. The historical record combines displacement from established activities with the formation of activities that earlier classifications could not name (Mokyr, Vickers, and Ziebarth, 2015; Autor, 2015). Employment did not simply disappear. It moved, often painfully, into different work.

This paper studies a recurring feature of technological change: employment adjusts inside established activities while successive entrants create additional activities and paid work. These two margins need not move together. The computing era provides a clear accounting example. Firms entering each year contributed large positive job flows even when incumbent net flows were weak or negative. Over time, the payroll retained by successive entrant cohorts accumulated into a substantial share of employment.

This paper makes four contributions. First, it documents the two margins in the computing era with Census firm-age data: recurring entry supplied all net job creation while incumbents contracted on net, successive entrant cohorts accumulated into two-fifths of payroll, and, across industries, those most exposed to automation saw incumbents fall behind from the early 1980s and entrants pull ahead later. The industry evidence is precise for the early incumbent shortfall and suggestive for the later entrant build-up, whose standard errors exceed the estimates by 1998. Second, it builds a model in which producers automate when compute, review, and correction deliver an acceptable result more cheaply than human production, entrepreneurs invest in products that still require development and commercialization, and operating products take time to re-scale. The model separates founder effort from hired hours and reports real wages, consumption, and active products alongside labor demand. It shows that aggregate displacement requires the scale friction: a sudden efficiency gain cuts hired hours by 4.6 percent at once, while the same gain diffusing over seven years, as adoption has, produces a dip of a quarter of a percent that closes after about seven years. Third, it carries the same industry design into the AI era with establishment job flows through the first quarter of 2026 and finds no gradient on language-model exposure on either margin; the model predicts exactly that under the observed adoption path, and so would the absence of any payroll effect, so the entry evidence to date rests on applications and registrations, where recent studies find formation rising 10 to 20 percent in founding-task-compatible industries. Fourth, it proposes how to measure the creation margin before it reaches payroll and shows how measurement lags distort the record.

Section 2 presents the historical evidence. Section 3 sets out the framework and its extension. Section 4 places AI on the clock of earlier general-purpose technologies and reports the simulations. Section 5 follows the two margins in early AI-era data. Sections 6 and 7 turn to measurement, and Section 8 draws implications.

\section{What past revolutions teach}

\subsection{Complementary investment and the timing of expansion}

The first industrial revolution combined mechanization of established production with the expansion of factory, engineering, transport, and commercial activity. Its gains did not accrue uniformly to workers or arrive at the same date as the first machines. The relevant historical lesson is that technical capability, commercial scale, and labor-market adjustment are distinct stages (Mokyr, Vickers, and Ziebarth, 2015; Autor, 2015).

Electrification made this distinction especially visible. Electric motors gradually replaced factory drive capacity, while firms redesigned layouts, transmission systems, and production processes around the new technology. Figure~\ref{fig:1} places that diffusion beside the later accumulation of information equipment and software. Computing similarly transformed routine and nonroutine work through decades of investment, integration, and organizational change (David, 1990; Devine, 1983; Autor, Levy, and Murnane, 2003).

The general-purpose-technology literature explains why complements matter. Bresnahan and Trajtenberg (1995) emphasize innovation complementarities across application sectors. Helpman and Trajtenberg (1998) formalize investment before the broader gains arrive. Brynjolfsson, Rock, and Syverson (2021) show how complementary intangible investment changes the timing of measured productivity. These mechanisms motivate examining creation and adjustment separately, rather than judging a transition from a single incumbent-employment series.

\subsection{The computing era: annual creation and incumbent adjustment}

Where did employment growth come from as computing spread: expansion within established firms, or the arrival of new employers? Aggregate employment cannot distinguish these contributions. We use the Census Business Dynamics Statistics (BDS) to separate jobs created at entering firms from net changes among firms already operating. This comparison establishes the importance of entry before asking how much employment successive entrants retain.

Figure~\ref{fig:2} reports annual net job creation during 1982-1998. In the BDS 2023 release, entrants at age zero added an average 2.78 million payroll jobs per year. Young incumbents aged one to five contributed -0.44 million net jobs, and older incumbents -0.52 million. The independently published national total averaged 1.80 million; Table~\ref{tab:A1} reports the components and publication residual. Recurring entry was large enough to exceed the combined net contraction among incumbents.

These are annual flows. A firm appears among entrants when it first becomes an employer and among incumbents in later years. Entry brings previously unrecorded payroll into the employer base; incumbent flows combine expansion, contraction, and closures. The graph therefore shows how the formation of new employers contributed to aggregate job creation alongside adjustment within existing firms.

\subsection{The common mechanism}

Acemoglu and Restrepo (2019) formalize what these episodes share. Technology has a displacement effect, as machines take over tasks previously done by people, and a reinstatement effect, as new tasks are created in which labor has a comparative advantage. Long-run employment depends on the balance between the two. The empirical weight of reinstatement is large. Using eight decades of Census job titles, Autor et al. (2024) estimate that roughly 60 percent of US employment in 2018 was in job titles that did not exist in 1940; among professionals the share is 74 percent. Their replication data also give the flow of new work decade by decade, which can be read against the computing clock.

Figure~\ref{fig:3} labels each bar by the year in which employment is measured. In 1990, 3.8 percent of labor input is allocated to titles added during the 1980s; in 2000, 11.4 percent is allocated to titles added during the 1990s. The 2000 bar is positioned 23 years after the first mass-market PCs in 1977. These are period-specific employment shares, not cumulative totals or growth rates. In 2018, titles added during 2000-2018 account for 9.9 percent of labor input. The electricity-era observations are 9.7 percent in 1940 for titles added during the 1930s and 26.7 percent in 1950 for titles added during the 1940s; the latter also reflects index reorganization. Each observation uses its own end-period employment denominator.

Lin (2011), using a different catalog baseline for each decade, puts employment in new work at 8.5 percent in 1980, 8.2 percent in 1990, and 4.4 percent in 2000, and finds that new work concentrates in dense, skill-intensive places. The historical record also carries a warning: adjustment can take a generation, and its distributional costs can be politically destabilizing (Frey, 2019). Task creation and employer entry are complementary ways to observe expansion: new tasks can arise inside existing firms, while new firms can employ workers in familiar occupations.

\subsection{Successive entrants and the employment they retain}

The annual flow comparison leaves an important question: does entry add lasting employment, or do those jobs soon disappear? To answer it, we divide firms into two groups according to when they first appeared as employers: those already operating before 1978 and those entering in 1978 or later. A firm stays in the same group throughout the comparison. For example, a business that entered in 1980 still counts toward the later-entry group in 1998, even though it is then 18 years old. This follows the employment accumulated by successive entrants as they mature, rather than counting only businesses that are currently young.

Figure~\ref{fig:4} and Table~\ref{tab:A2} show the difference. Employment in firms entering in 1978 or later rose from 2.55 million in 1978 to 45.68 million in 1998, a gain of 43.13 million. Employment in firms already operating before 1978 fell by 5.69 million. Together, these changes account for 37.43 million additional payroll jobs. By 1998, the later-entry group supplied 42.4 percent of payroll employment. Entry contributed not only an annual inflow, but also a growing stock of employment that remained as those firms aged. The earlier group adds no new firms, but its employment can rise when expansion at surviving firms exceeds contraction and exits. The later group combines the arrival of new employers with their subsequent growth, contraction, and exit.

The published firm-age categories identify the two groups exactly at dates when the age-band boundary coincides with the 1978 entry cutoff. At intervening dates, Figure~\ref{fig:4} shows the range consistent with the band that spans the cutoff. \hyperref[app:B]{Appendix B} describes the construction. This preserves the same entry-year classification throughout the comparison.

Table~\ref{tab:A3} complements this stock comparison with annual net job-creation rates by firm age. It reports ages zero through five individually, ages six to ten, and ages eleven to fifteen, with firms aged sixteen and above as the reference. The sample is 1992-1998, when the pre-1977 left-censored firms can all be assigned to the oldest group. The four specifications add macroeconomic controls and sector fixed effects separately and together. In the combined specification, the age-two rate is 3.79 percentage points below the reference, the ages-six-to-ten rate is 1.21 points below, and the ages-eleven-to-fifteen rate is 0.35 points below the reference and is not statistically distinguishable from it. The approximately 200-point age-zero contrast reflects the symmetric growth-rate normalization for entrants; Figure~\ref{fig:2} shows their contribution in numbers of jobs.

Separately, Figure~\ref{fig:4} and Table~\ref{tab:A4} follow the same entering cohorts through their first five years across 19 sectors. Entering cohorts retain about 80 percent of initial payroll after five years, with a 95 percent interval of 77.35-82.63 percent across cohorts. With macroeconomic controls and sector and birth-cohort fixed effects, the age-five difference is -18.93 percentage points relative to entry (standard error 1.17). Table~\ref{tab:A4} reports the baseline comparisons and robustness checks. Repeated entry can therefore contribute substantial cumulative employment even while a particular cohort contracts.

Taken together, the annual flows and cohort comparisons show why both margins matter. Established firms adjust their workforces, while successive entrants add employment that can persist well beyond their first year. The aggregate contribution of entry can be substantial even when an individual entering cohort subsequently contracts. These historical trends during the computing era are consistent with the model's two margins: adjustment within existing production and employment associated with new activity.

\subsection{The two margins by industry automation-patent exposure}

The three-digit BDS industry panel follows the two margins annually in 84 industries across 19 sectors from 1978 to 1998. For this annual comparison, the original employer base is the pre-1977 left-censored group; later origins begin in 1977. Unlike the 1978 cutoff in Table~\ref{tab:A2}, this split is exact every year. Later origins held more than one-quarter of 1998 employment in 66 of the 84 industries and more than two-fifths in 43. Entry and subsequent growth were broad features of the economy rather than confined to technology-producing industries.

Figure~\ref{fig:5} compares employment changes at firms operating before 1977 and firms entering in 1977 or later across industries with different automation-patent exposure. Exposure combines 1980 occupational composition with the occupation-level 1980s patent measures of Autor et al. (2024). Every curve uses total 1978 industry employment as its denominator, so the two firm-group contributions add to the total. The solid line is the employment-weighted industry mean. The other lines show fitted comparisons at exposure one standard deviation higher or lower, holding initial size and sector effects fixed and anchoring both comparisons to that mean. They are not averages of observed high- and low-exposure groups. The exposure is retrospective, not a treatment fixed before 1978.

The automation-patent gradient is most apparent initially in the original employer base. By 1982, one standard deviation of exposure is associated with 4.95 percentage points less incumbent employment relative to the common 1978 denominator (standard error 1.84), versus 0.82 points more employment at later origins (standard error 1.28). By 1998 these gradients are -9.39 and +5.94 points, with standard errors 6.57 and 7.17. The point estimates show an early incumbent shortfall alongside a later build-up of the entrant contribution. The incumbent shortfall is precisely estimated through the 1980s; by 1998 both standard errors exceed the estimates, so the later build-up and the offset fraction are suggestive rather than precise. Routine-cognitive intensity yields small, imprecise gradients on both margins. Tables A5-A8 report specifications, annual estimates, and robustness; \hyperref[app:E]{Appendix E} details the construction.

\section{A framework for the two margins}

Consider two uses of the same AI capability. An established software business uses AI to resolve support tickets. It already has a product, customers, a support queue, and standards for judging successful resolution. Once integration and quality requirements are met, AI can reduce human minutes per accepted answer. An entrepreneur uses AI to build a software product. Coding may become cheaper immediately, but a working prototype is not yet a business. Demand validation, reliability, approvals, customer acquisition, and distribution may still be unfinished.

The hypothesis concerns these different initial conditions. AI can improve established activities using infrastructure already in place, while additional products expand through investment and completion. During experimentation, founders may supply the effort themselves, or an established firm may reassign staff. As products pass validation, find customers, and expand, firms can hire workers for development, delivery, and operations. The framework captures that progression by separating the cost of operating existing products from the investment needed to create new ones, and founder time from hired labor.

\subsection{Households, hired hours, and founder effort}

Time is continuous. A representative household owns active producers, unfinished projects, and compute suppliers. It supplies hired hours H and own-account founder effort F. The latter is initiation and development time that is not paid through an employment relationship. Both use scarce time. Let w be the real wage and \(w_F\) the shadow opportunity cost of founder time, valuing its marginal time cost. Preferences are:

\begin{equation}\label{eq:preferences}
\begin{gathered}
\int_0^\infty e^{-\rho t}\left[u_\sigma(C_t)-\frac{\chi_H H_t^{1+1/\phi}}{1+1/\phi}-\frac{\chi_F F_t^{1+1/\phi_F}}{1+1/\phi_F}\right]dt, \\
u_\sigma^{\prime}(C)=C^{-\sigma}.
\end{gathered}
\end{equation}

The household's first-order conditions equate the marginal cost of time to its return. Hired wages are cash compensation; \(w_F\) is an opportunity cost, not an imputed payroll payment:

\begin{equation}\label{eq:hours}
w_t C_t^{-\sigma}=\chi_H H_t^{1/\phi},\qquad w_{F,t}C_t^{-\sigma}=\chi_F F_t^{1/\phi_F}.
\end{equation}

The benchmark allows partial income effects, setting sigma=0.6. Log consumption utility, sigma=1, provides the stronger-income-effect sensitivity, while sigma=0 removes income effects. With phi=0.5, a 1 percent increase in consumption reduces supplied hired hours by approximately 0.30 percent at a fixed wage. Evidence from lottery winners supports an income effect of this order: Cesarini et al. (2017) find that Swedish winners reduce earnings persistently, with a marginal propensity to earn out of unearned income near -0.1, and Golosov et al. (2024) estimate a larger propensity, around -0.5, for US winners. The benchmark sits between the two cases that bracket this range. Greater purchasing power can therefore support more leisure. H is paid labor input, not worker headcount: the same hours can be supplied by different combinations of workers and hours per worker. Founder work paid through an employee salary belongs to hired hours, rather than the own-account category.

\subsection{Products and the compute constraint}

A competitive final-good sector combines M active varieties with CES substitution elasticity epsilon greater than one. Each variety has a unique producer, which may belong to a multiproduct firm. Let Y denote gross output and Q total physical product quantity. With the composite price normalized to one, the flexible-scale reference has symmetric quantities and monopolistic pricing:

\begin{equation}\label{eq:products}
\begin{gathered}
Y_t=\left[\int_0^{M_t}q_{jt}^{(\varepsilon-1)/\varepsilon}dj\right]^{\varepsilon/(\varepsilon-1)}, \\
p_t=M_t^{1/(\varepsilon-1)}=\frac{\varepsilon}{\varepsilon-1}c_t,\qquad Y_t=p_tQ_t.
\end{gathered}
\end{equation}

Compute uses final goods. A convex resource requirement generates an upward-sloping supply curve, with elasticity eta and infrastructure scale kappa. Consumption is output after paying the real resource cost of compute:

\begin{equation}\label{eq:compute}
\Psi_\eta(X)=\frac{\eta}{1+\eta}\kappa^{-1/\eta}X^{1+1/\eta},\quad p_X=\Psi_\eta^{\prime}(X),\quad X=\kappa p_X^\eta,\quad C+\Psi_\eta(X)=Y,\quad\eta>0.
\end{equation}

\subsection{Automation is a cost and quality decision}

One product unit requires a continuum of tasks of unit mass. Human production uses one hired hour per task. An AI-assisted method uses k/gamma units of compute and v hired hours of review and expected correction per acceptable result. The requirement k includes retries and has distribution \(G_k\). Both methods deliver the same accepted output. Firms choose the less expensive method:

\begin{equation}\label{eq:adoption}
c_t(k)=\min\left\{w_t,\frac{p_{X,t}k}{\gamma_t}+w_tv\right\},\qquad k_t^*=\frac{\gamma_t(1-v)w_t}{p_{X,t}},\qquad a_t=G_k(k_t^*).
\end{equation}

The cutoff is truncated to the support of task requirements. The resulting human and compute requirements per product unit are:

\begin{equation}\label{eq:inputs}
\ell_t=1-(1-v)a_t,\qquad x_t=\gamma_t^{-1}\int_{k\leq k_t^*}k\,dG_k(k),\qquad c_t=w_t\ell_t+p_{X,t}x_t.
\end{equation}

Proposition 1 (economic adoption). A producer assigns a task to AI exactly when its full acceptable-output cost does not exceed human cost. A higher compute price lowers the cutoff, while greater AI efficiency raises it. Compute-market clearing makes assignment respond to the infrastructure used by all producers. For established workflows, the benchmark takes deployment infrastructure as available. A fixed integration investment would additionally require discounted savings to cover that investment.

\subsection{Entrepreneurial development before operation}

Let P be unfinished projects and E new project starts. Each unfinished project uses founder effort e and hired development hours h. An effective development input d combines them. Its unit cost \(c_D\) prices founder time at its opportunity cost:

\begin{equation}\label{eq:development}
\begin{gathered}
d=\left(\frac{e}{\theta}\right)^\theta\left(\frac{h}{1-\theta}\right)^{1-\theta},\qquad c_D=w_F^\theta w^{1-\theta}, \\
e=\frac{\theta c_Dd}{w_F},\qquad h=\frac{(1-\theta)c_Dd}{w}.
\end{gathered}
\end{equation}

For theta equal to one, development uses founder effort alone; otherwise theta is the founder share of development cost. Completion occurs at intensity nu(d), projects are abandoned at rate omega, and active products exit at rate delta. Project initiation requires own-account effort I(E), with increasing marginal requirements:

\begin{equation}\label{eq:stocks}
\begin{gathered}
\nu(d)=\nu_0d^\xi,\quad 0<\xi<1; \\
\dot P_t=E_t-[\nu(d_t)+\omega]P_t,\qquad \dot M_t=\nu(d_t)P_t-\delta M_t.
\end{gathered}
\end{equation}

\begin{equation}\label{eq:initiation}
I(E)=bE+\frac{\zeta}{2}E^2.
\end{equation}

The convex term captures increasingly costly ideas or scarce complementary capabilities. Each active product also requires \(f_m\) hired maintenance hours and earns operating profit pi. Let V and J be the consumption-good values of an active product and an unfinished project, and let r be the household discount rate:

\begin{equation}\label{eq:profits}
\pi_t=\frac{p_tQ_t}{\varepsilon M_t}-w_tf_m,\qquad r_t=\rho+\sigma\frac{\dot C_t}{C_t}.
\end{equation}

\begin{equation}\label{eq:values}
\dot V_t=(r_t+\delta)V_t-\pi_t,\qquad \dot J_t=[r_t+\omega+\nu(d_t)]J_t+c_{D,t}d_t-\nu(d_t)V_t.
\end{equation}

Optimal effort trades the cost of development against faster completion; project starts equate project value to the marginal opportunity cost of initiation:

\begin{equation}\label{eq:entry}
c_{D,t}=\nu^{\prime}(d_t)(V_t-J_t),\qquad E_t=\max\left\{0,\frac{J_t/w_{F,t}-b}{\zeta}\right\}.
\end{equation}

AI can improve accepted operating output through gamma, reduce the linear effort needed to initiate a project through b, or improve development effectiveness through \(\nu_0\). These are separate margins. Entry is based on expected project value, rather than zero operating profit in every period. Existing producers start with customers and usable infrastructure; newly initiated projects start with unfinished requirements.

\subsection{Timing, labor accounting, and the offset condition}

Proposition 2 (new cohorts expand through completion). Consider an additional cohort with no initial unfinished or active projects. If its initial initiation flow is Delta E and completion intensity is bounded and initially nu, then for a short interval after an AI improvement:

\begin{equation}\label{eq:timing}
\Delta P_t=\Delta E\,t+o(t),\qquad \Delta M_t=\frac{1}{2}\nu\Delta E\,t^2+o(t^2).
\end{equation}

The cohort's operating scale accumulates after project starts. With founder-only development, its prelaunch effort creates no hired development hours; with hired development, that demand grows as the project stock builds. Existing unfinished projects can finish sooner when effort increases, so the initial pipeline changes the aggregate response.

The resource accounting distinguishes paid work from own-account effort at every date:

\begin{equation}\label{eq:accounting}
H_t=\ell_tQ_t+f_mM_t+h_tP_t,\qquad F_t=e_tP_t+I(E_t),\qquad L_t=H_t+F_t.
\end{equation}

Proposition 3 (offsetting the reduction in hired hours). Relative to the initial equilibrium, hired hours are at least as high exactly when operating-scale expansion, product maintenance, and hired development cover the direct task-level saving:

\begin{equation}\label{eq:offset}
\ell_t(Q_t-Q_0)+f_m(M_t-M_0)+(h_tP_t-h_0P_0)\ \geq\ Q_0(\ell_0-\ell_t).
\end{equation}

Founder effort is not added to the left-hand side as though it were paid employment. Infrastructure affects both sides through acceptable-output costs, profits, investment, and completion. An equilibrium satisfies these allocation identities, household optimality, task assignment, pricing, the project-stock laws, and compute and goods-market clearing. Values respond to expected returns while the stocks of projects and operating products are predetermined. \hyperref[app:A]{Appendix A} gives the proofs and numerical solution.

\subsection{Extension: operating products take time to scale}

Existing products can deploy a cheaper method without immediately selling more output. Customer acquisition, distribution, installed capacity, and process redesign may still need to adjust. This is a second timing margin, alongside the development of new products. We represent it by a reduced-form quantity-adjustment law for products operating when AI improves, consistent with the complementary-organization mechanism discussed by David (1990). Later products choose their operating scale flexibly. The household, task-assignment, development, and compute blocks are unchanged.

\begin{equation}\label{eq:scale_vintages}
\begin{gathered}
M_t^I=M_0e^{-\delta t},\qquad M_t^N=M_t-M_t^I, \\
\dot q_t^I=\lambda(q_t^N-q_t^I),\qquad q_0^I=q_0.
\end{gathered}
\end{equation}

The inherited quantity is predetermined, while the later-product quantity is the flexible target. The speed \(\lambda\) is an adjustment parameter, not an optimized investment choice. At \(\lambda\)=0 inherited quantity stays fixed until the product exits. With an unchanged target, \(\lambda\)=0.2 per year gives a 3.47-year half-life; the realized recovery time also depends on wages, demand, and entry. The flexible-scale reference removes this constraint and imposes equal quantities, the infinite-speed limit.

\begin{equation}\label{eq:scale_ces}
\begin{gathered}
Y_t=[M_t^I(q_t^I)^{(\varepsilon-1)/\varepsilon}+M_t^N(q_t^N)^{(\varepsilon-1)/\varepsilon}]^{\varepsilon/(\varepsilon-1)}, \\
Q_t=M_t^Iq_t^I+M_t^Nq_t^N.
\end{gathered}
\end{equation}

\begin{equation}\label{eq:scale_prices}
p_t^j=(q_t^j/Y_t)^{-1/\varepsilon},\quad j\in\{I,N\},\qquad p_t^N=\frac{\varepsilon}{\varepsilon-1}c_t.
\end{equation}

CES demand determines the price of each inherited quantity. Later products charge the markup price. These conditions replace symmetric pricing during the transition; both vintages face the same unit input cost. Compute clearing remains X=xQ and consumption is output net of compute resources. Incumbent profits accrue to the household, while development and initiation value a newly completed, flexibly scaled product:

\begin{equation}\label{eq:scale_profits}
\pi_t^N=\frac{p_t^Nq_t^N}{\varepsilon}-w_tf_m,\qquad \pi_t^I=(p_t^I-c_t)q_t^I-w_tf_m.
\end{equation}

The value recursions apply with V as the value of a later product and its operating profit. Propositions 1-3 continue to hold: task assignment is cost based, new cohorts accumulate through completion, and the hired-hours offset condition is an accounting identity. For finite \(\lambda\), physical output is fixed on impact, but hired development can respond immediately. The impact change in hours includes both task savings and the change in hired development. \hyperref[app:D]{Appendix D} gives the exact identity and solution.

\section{The AI parallel}

\subsection{Four technologies on one clock}

AI qualifies as a general purpose technology in the sense of Bresnahan and Trajtenberg (1995): pervasive, improving, and a spur to complementary innovation. Eloundou et al. (2024) estimate that about 80 percent of US workers could see at least 10 percent of their tasks affected by large language models, and about 19 percent could see at least half affected. The exposure is broad and, unlike earlier waves, reaches deep into cognitive and professional work.

To compare transitions we align each on the same stage. Every general purpose technology here passed through three: a breakthrough (Edison's lamp and the Pearl Street station in 1879--82; the microprocessor in 1971; AlexNet in November 2012), a first usable product (the Niagara transmission line in 1896; the Apple II, Commodore PET and TRS-80 in 1977; the commercial internet in 1995; GPT-2 in February 2019), and a mass standard (the IBM PC in 1981; ChatGPT in November 2022). We count years from the first usable product and mark the mass-standard moment, which came 3.7 years after GPT-2 and 4 years after the first PCs. The lag from breakthrough to product has shortened: fourteen years for electricity, six for computing and for AI; from product to standard it has stayed at about four. On this clock, diffusion at work has been fast but not without precedent. The Census Bureau asked employed adults from 1984 whether they used a computer at work: 25 percent did in 1984, seven years after the first PCs; 37 percent in 1989; 46 percent in 1993; and 56 percent in 2003 (US Census Bureau, CPS; BLS, 2005). The same question about the internet gives 16 percent of the employed in 1997, two to three years after the commercial opening, 26 percent in 2000 and 42 percent in 2001 and 2003 (NTIA, 2002; BLS, 2005). The question for generative AI, asked quarterly by Bick, Blandin and Deming's tracker, gives 33 percent of employed adults using it for their job in mid-2024 and 45 percent by the second quarter of 2026, seven years after GPT-2: the level the internet reached at about the same age and the PC did not reach for sixteen years (Figure~\ref{fig:6}). At the level of firms, 24 percent of US employer businesses were using AI by October 2026, close to where computer use by workers stood seven years after the first PCs; in production, electric motors supplied a tenth of factory power eight years after Niagara.

Adoption at home tells the same story: 39 to 45 percent of adults were using generative AI at years five and six (Bick, Blandin and Deming, 2026), the share households with internet access reached at year five and PC ownership and electric service did not reach until years 22 and 26 (US Census Bureau, CPS; Historical Statistics, S108). Firm adoption lags use by workers, as it did for electricity and the PC. The Census Bureau's Business Trends and Outlook Survey found 3.7 percent of employer businesses using AI to produce goods or services in September 2023 and 10 percent by October 2025; under a broader question introduced in November 2025, covering AI in any business function, the share was 17 percent at the outset and 24 percent by October 2026 (US Census Bureau, 2026).

\begin{figure}[H]
\centering
\includegraphics[width=.49\textwidth]{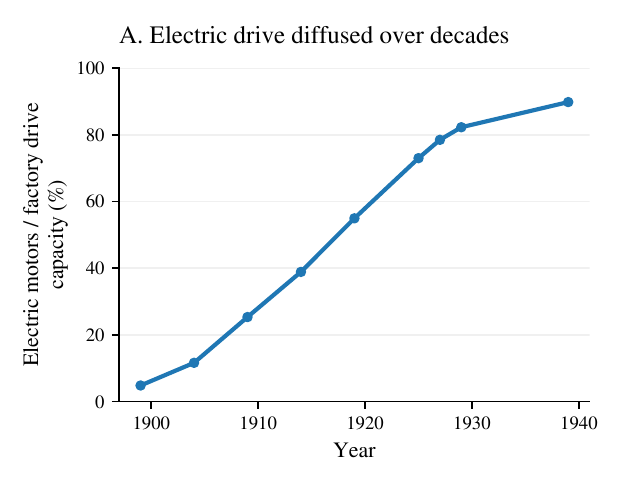}\hfill
\includegraphics[width=.49\textwidth]{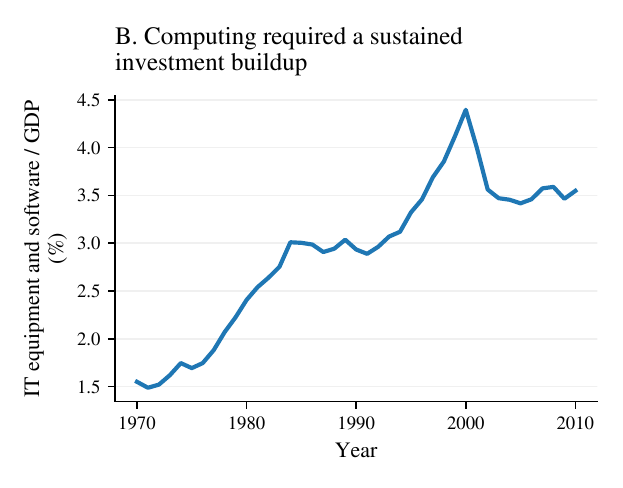}\\[7pt]
\caption{Complementary infrastructure in past revolutions}\label{fig:1}
\begin{minipage}{.96\textwidth}\footnotesize Notes. Sources: Historical Statistics of the United States manufacturing drive-capacity series Dd848-853 and Devine (1983); BEA information-equipment and software investment and GDP in the input workbook. Panel A uses the comparable factory basis from 1899. Panel B is a nominal investment share of GDP, 1970-2010. These panels retain their own historical units and do not impose a common treatment date.\end{minipage}
\end{figure}

\begin{figure}[H]
\centering
\includegraphics[width=.96\textwidth]{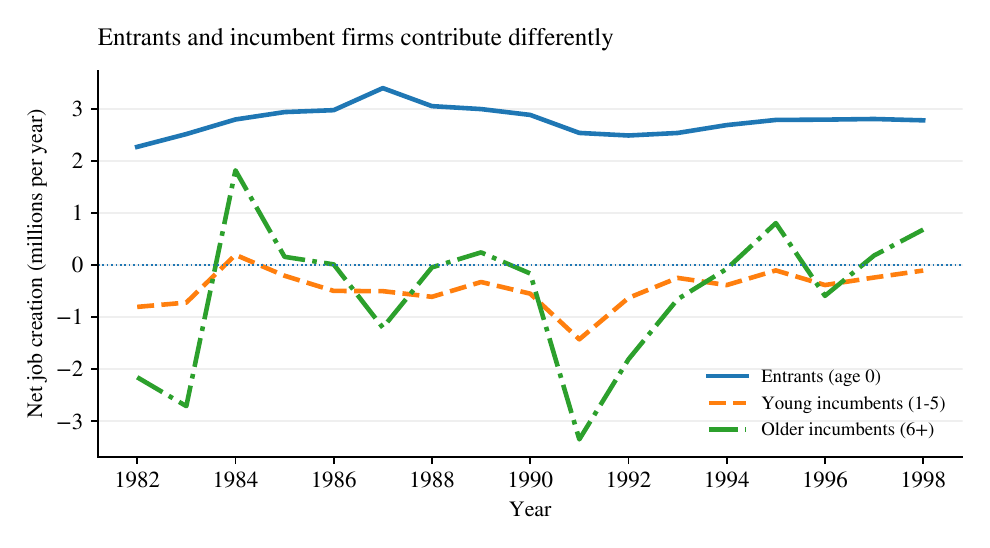}\\[5pt]
\caption{Entry and incumbent adjustment during the computing era, 1982-1998}\label{fig:2}
\begin{minipage}{.96\textwidth}\footnotesize Notes. Source: Census BDS, 2023 release, bds2023\_pfg\_utly\_fa.csv, summed across its mutually exclusive patent statuses. Annual net payroll-job creation by contemporaneous firm age, in millions. Entrants are age zero, young incumbents ages one to five, and older incumbents ages six and above, including the left-censored group. Firms move between groups as they age. Table~\ref{tab:A1} reports levels and the residual against the separate national release. These flow categories are different from the fixed-origin stock groups in Figure~\ref{fig:4}.\end{minipage}
\end{figure}

Complementary investment is expanding rapidly. US private AI investment reached \$109 billion in 2024 and \$286 billion in 2025 (Stanford AI Index, 2025, 2026). The historical comparison in Figure~\ref{fig:1} likewise shows sustained investment before widespread productivity gains. These series measure different assets and prices, so their growth rates are not a common estimate of the speed of technological transformation. The AI-era dashboard separates investment, emerging roles, and the broader business-formation environment.

Figure~\ref{fig:6} compares the four technologies on this common clock. Figure~\ref{fig:7} then places the current computer-services payroll series beside the historical decline in the employment share of clerical and administrative occupations, keeping the two displacement comparisons on their own measurement scales.

On employment, early evidence is mixed in a way that echoes previous transitions. Using Danish administrative data, Humlum and Vestergaard (2025) estimate precise null effects on earnings and recorded hours, ruling out effects larger than 2 percent two years after adoption. Brynjolfsson et al. (2025) find a 16 percent relative employment decline for workers aged 22 to 25 in the most AI-exposed occupations, concentrated where AI automates rather than augments. Revelio Labs' monthly tracker, which replicates that design on profile data, put the gap at 20 percent for the youngest workers and 7 percent for all workers in the most exposed occupations by August 2026, with job postings in those occupations 29 percent below the least exposed; at the same time, headcount in AI-related roles had grown 19 percent since November 2022 against 3 percent for all other roles, and firms that had adopted AI were growing headcount faster than non-adopters (Revelio Labs, 2026). A study of nearly all US job postings from 2019 to March 2025 finds postings for routine, automation-prone roles down 13 percent after ChatGPT and postings for analytical, technical and creative roles up 20 percent (Chen, Srinivasan and Zakerinia, 2026).

Two distinctions matter for the posting-based evidence. Postings are not hires: recruitment also occurs through referrals, internal moves, and contractors, and displaced workers can move into non-payroll activity. Private trackers are anchored to their own coverage and comparison dates. The public computer-services payroll series in Figure~\ref{fig:7} also declines after its late-2022 peak. Together, these observations identify adjustment in established work. The creation side is followed separately through project formation, investment, and subsequent hiring.

\subsection{The flexible-scale reference: two labor-demand margins}

AI acts on both historical margins. It reduces the cost of acceptable output within existing production, and it makes additional projects worth attempting. In the numerical experiment, compute efficiency permanently rises by 50 percent and the linear founder effort required to initiate a project falls by 25 percent at time zero. Both gains become available together. Development and completion determine when the second margin becomes active production; no calendar delay is imposed on entry.

Table~\ref{tab:A9} gives the calibration. We use partial income effects, sigma=0.6, as the single benchmark. Product substitution is 4, implying a 33 percent markup, and hired-hours responsiveness is 0.5, guided by Bilbiie, Ghironi, and Melitz (2012) and Chetty et al. (2013). Compute supply has elasticity one. Founders initially provide 80 percent of development cost, and initiation uses founder time. Task heterogeneity, review requirements, completion rates, founder responsiveness, and shock sizes define the numerical experiment. Table~\ref{tab:A10} varies these inputs around the same benchmark and separately varies household income effects.

Figure~\ref{fig:8} divides the response into hired hours attached to products already operating at the shock and hired hours attached to products that subsequently become active, including hired development. Each is measured against the corresponding no-shock path, so normal entry and exit are not misclassified as AI-induced creation. Founder effort is reported separately. This decomposition mirrors the distinction between an existing employer base and subsequent entry without identifying a product variety with a legal firm.

The industry comparison in Figure~\ref{fig:5} provides a descriptive counterpart to this distinction: employment at an inherited employer base can move differently from employment accumulated by later origins. Here the two product contributions are measured relative to model no-shock paths, rather than across observed industries.

\begin{equation}\label{eq:legacy_decomp}
\begin{gathered}
M_t^I=M_0e^{-\delta t},\qquad g_t=\ell_tQ_t/M_t+f_m, \\
\Delta H_t^I=M_t^I(g_t-g_0),\qquad \Delta H_t^N=\Delta H_t-\Delta H_t^I.
\end{gathered}
\end{equation}

At year 15 in the benchmark, products operating before the shock contribute -2.16 percent of initial hired hours relative to their no-shock path, while later products and hired development contribute 2.42 percent. Their sum is a 0.27 percent increase in aggregate hired hours. Across the five reported preference cases, the year-15 existing-product contribution is negative, the new-activity contribution positive, and real wages and consumption higher.

In the sigma=0.6 benchmark, hired hours fall 0.07 percent on impact and return to their no-shock level after about 4.1 years. The initial drop followed by gradual recovery forms a modest J-shaped response; the trough occurs immediately at the shock. The aggregate balance varies with preferences. With no income effect, output expansion raises hired hours even in existing products on impact, and total hours are 2.31 percent higher at year 15. With sigma=1, total hours remain 0.57 percent below baseline at year 15 while consumption rises. Figure~\ref{fig:8} compares these paths under the same technology shock.

Figure~\ref{fig:9} uses the same sigma=0.6 benchmark. By year 15, real wages rise 4.41 percent, consumption 6.51 percent, and active products 5.03 percent; founder effort rises 0.69 percent. With compute-supply elasticity 0.25 rather than 4, consumption rises 5.70 rather than 8.03 percent. Higher purchasing power and more viable products accompany a modest increase in hired hours in the benchmark. Founder effort remains a separate resource, not an imputed payroll payment.

\subsection{Incumbent-scale adjustment and an efficiency gain diffusing over seven years}

We retain sigma=0.6, the 50 percent compute-efficiency improvement, and the 25 percent reduction in linear initiation effort. The extension sets \(\lambda\)=0.2 per year and keeps the initial calibration targets unchanged. Figure~\ref{fig:10} shows a sudden efficiency improvement; Figure~\ref{fig:11} shows the alternative in which the efficiency gain diffuses over seven years. Table~\ref{tab:A11} reports impact and recovery; Table~\ref{tab:A12} reports year-15 outcomes. All cases retain flexible wages and market clearing.

With a sudden improvement and finite scaling, hired hours fall 4.58 percent on impact, compared with 0.07 percent in the flexible-scale reference. Task savings reduce hours by 4.75 percent of initial hired hours, while additional hired development contributes 0.17 percentage points. The net decline is therefore smaller than the operating-task saving alone. Real wages initially fall 9.23 percent as labor supply and demand adjust, raising the flexible output target and accelerating incumbent rescaling. Aggregate hired hours recover around year 3.0 and reach 0.32 percent above baseline at year 15.

The quantity-adjustment speed and the aggregate recovery date are different objects. With no inherited rescaling, later products and product turnover still restore hired hours around year 11.1. At \(\lambda\)=0.2, the sigma=1 sensitivity remains 0.52 percent below baseline at year 15. The income-effect sensitivity therefore continues to govern whether higher living standards come with more or fewer hired hours.

In the diffusing case, effective AI efficiency phases in along the normalized logistic path in \hyperref[app:D]{Appendix D}, while the initiation-cost improvement is available immediately. Task adoption remains an endogenous cost comparison. Efficiency reaches half of its improvement after 3.62 years and 95 percent after 7.10 years. With \(\lambda\)=0.2, hired hours reach a trough 0.23 percent below baseline near year 3.8, recover around year 6.8, and rise 0.32 percent by year 15. The real wage has a small dip of 0.10 percent. The productivity path changes the depth and timing of adjustment without imposing a calendar date for employment recovery.

Which combinations of adjustment speed and diffusion are consistent with the absence of a decline in payroll? Table~\ref{tab:A13} reports the aggregate trough and the recovery date for \(\lambda\) from 0 to infinity and for efficiency gains that arrive at once or diffuse over 2 to 10 years. A sudden gain produces a trough of 4.5 to 4.6 percent for every finite \(\lambda\) and is excluded by the data in Section 5. With diffusion over two years, troughs of 1.1 to 1.9 percent at \(\lambda\) of 0.2 or below would have appeared as a stall in aggregate payroll. With diffusion over five years or more no combination exceeds 1 percent, and at the seven-year pace of Figure~\ref{fig:6} none exceeds 0.6 percent. The data reject sudden arrival, bound \(\lambda\) only if diffusion was fast, and under the observed pace leave \(\lambda\) unidentified: the two margins offset each other within the resolution of payroll statistics.

\section{Early AI indicators along the two margins}

\subsection{Formation and the expansion of activity}

Studies that measure exposure at the level of founding tasks find that generative AI has raised entry. Marchesi and Tang (2026) build an industry-level measure of how far AI can support the tasks involved in starting a firm and estimate that, after the release of ChatGPT, firm formation rose by more than 10 percent in the more compatible industries; the increase is concentrated where AI can assist and experimentation is critical, and the new firms are more likely to survive and grow. Bena, Bian, and Giannetti (2026) find about 20 percent more startup formation in industries with higher task-level exposure, with smaller entrants, founders drawn from high-exposure occupations, entry extending beyond venture hubs, and net employment and wage growth at the industry level, including in the occupations most exposed. At Census, Dinlersoz, Dogan, and Zolas (2024) identify applications that aim to develop or use AI from the application text and find a discrete jump in 2023 and a higher likelihood of becoming employer businesses than other applications.

These results concern applications, platform-recorded formations, and venture counts. Payroll at opening establishments is a later stage, and it does not yet move with exposure. In BLS establishment data through the first quarter of 2026, industries whose workforces are more exposed to large language models show no more job gains at openings and no larger losses at continuing establishments than their sector peers since late 2022 (Figure~\ref{fig:12} and Table~\ref{tab:A14}; Appendix F). Applications responding before payroll is the sequence of Section 3.

This margin includes entrepreneurial activity outside payroll. A founder can explore several products using the same hours, and a successful business can remain founder-operated; the new evidence points to smaller entrants. On freelance platforms, AI-related projects carry larger budgets and more hours than comparable work rather than higher hourly rates, and tasks that AI could perform continue to be purchased as paid work (OpenAI and Upwork, 2026). Applications, product launches, nonemployer revenue, founder time, and first hires therefore locate different stages of expansion. The historical entry evidence motivates observing those stages rather than treating the absence of immediate hiring as the absence of additional economic activity.

\subsection{Adjustment in established work}

Figure~\ref{fig:7} locates adjustment within existing activities in the two cases where it is most visible: clerical and administrative occupations lost employment share gradually during the computing era, and payroll in computer systems design, the industry most directly exposed to large language models, has declined from its late-2022 peak. Figure 12B widens the view to all continuing establishments. Across 72 industries, employment at establishments operating before ChatGPT fell by 13.6 percent of 2019 employment between 2022Q4 and 2026Q1 through contractions and closings net of expansions, at a quarterly pace no faster than over 2019-2022, and openings more than replaced it. Industries one standard deviation more exposed to large language models lost 14.7 percent and industries one standard deviation less exposed 12.5 percent, a difference of 1.1 points with a standard error of 1.9 (Table~\ref{tab:A14}). Measured in industry payroll, the incumbent margin has not yet moved with exposure.

In the AI transition, this margin can appear as lower human time per accepted output, a different task mix, slower hiring, reduced hours, or separations. Output growth and reassignment can offset task savings within an employer, so task content and hours provide additional information alongside job counts.

Through the first quarter of 2026, then, neither aggregate payroll nor the industry cross-section shows a decline. In the model this is the case of an efficiency gain diffusing over seven years: with efficiency diffusing on the path in Figure~\ref{fig:6}, aggregate hired hours reach a trough 0.23 percent below baseline and the reduction at operating products accumulates over about five years, within the sampling error of Figure~\ref{fig:12} at this horizon. A sudden improvement of the benchmark size would have reduced aggregate hired hours by 4.6 percent within a year; the aggregate series shows no such break, and Figure~\ref{fig:12} shows no cross-industry counterpart (Table~\ref{tab:A13} gives the combinations of adjustment speed and diffusion the data admit). The null is also what no effect on payroll would look like. At this horizon the diffusing case and the absence of an effect are observationally equivalent in payroll data, which is the measurement problem Section 6 describes; what separates them is the prediction that the difference appears first within occupations and in the hiring of entrants to exposed jobs, as Brynjolfsson et al. (2025) report, before it appears in industry payroll totals. The aggregate series bounds the decline relative to trend only loosely, since the same window contains the pandemic recovery and later shocks.

\section{The warning: our units of assessment may be inadequate}

If new activity is an important channel through which technology sustains labor demand, then our ability to see its formation becomes decisive. Five limitations stand out. First, occupational taxonomies lag reality. New activities can appear under residual categories or familiar professional titles before classifications change. Autor et al. (2024) recover new work by comparing successive job-title catalogs; a title-based measure is necessarily sensitive to the timing and scope of those catalog updates.

Second, the job is the wrong grain for some changes. AI acts on tasks. A lawyer, analyst, or engineer may keep the same title while the content of the role changes substantially. Occupational employment can remain unchanged when the worker shifts from producing a routine output to evaluating AI output or developing a new service. New tasks within an existing job and additional jobs are separate margins.

Third, activity can develop outside payroll. Founders can explore products using their own time, and independent professionals can supply project-based work without becoming employees. Establishment payroll surveys and household surveys cover these arrangements differently (Abraham et al., 2018). Founders' experimentation is evidence of an expanding set of feasible activities; it becomes evidence of additional paid employment only when the relevant employment relationship is observed. Even total effort need not rise if AI allows more projects to be attempted in the same working week.

Fourth, measured output does not fully capture economic value. Free or near-free digital services can generate substantial consumer benefits beyond the transactions recorded in national accounts (Brynjolfsson, Collis, and Eggers, 2019). A service can become more useful without a proportionate increase in spending or recorded output. Assessing AI's economic contribution therefore requires distinguishing measured production from the value users receive.

Fifth, timing shapes what appears in the data. A business application, prototype, first sale, and first hire can occur at different dates. An early payroll observation describes one stage of this process, not its complete outcome.

\section{Measuring the uptake}

The uptake has to be measured where it happens: in tasks, titles, and hours. We propose a core statistic and supporting indicators, and report a historical accounting analogue for the computing era. The analogue anchors measurement design; it is not an estimate of the AI reinstatement ratio.

Core statistic: the net reinstatement ratio. Following the displacement and reinstatement logic of Acemoglu and Restrepo (2019), compare paid hours in additional activities with the paid task hours displaced from established production since a common base year:

\begin{equation}\label{eq:uptake}
\mathcal{R}_t^{\mathrm{paid}}=\frac{\sum_j H_{jt}^{\mathrm{additional\ activities}}}{\sum_j H_{jt}^{\mathrm{displaced\ tasks}}}.
\end{equation}

A ratio above one means the additional activities use more paid hours than automation removes. Job-posting text, task records, and title data can locate new activities; displaced hours require a counterfactual for the work that would otherwise have been performed. Founder effort is measured alongside the paid-hours ratio. This distinguishes the exploration of new products from their eventual demand for hired workers.

A historical accounting benchmark. In the Autor et al. (2024) panel for 1990-2018, allocating 2018 labor input in proportion to newly added titles yields 20.1 percent in new work. Dividing this amount by losses in occupations that shrank relative to aggregate employment growth gives 1.09; using absolute losses instead gives 2.92. Thus imputed labor in new titles roughly matched relative losses elsewhere. \hyperref[app:C]{Appendix C} gives the construction. The 1.09 value is an accounting analogue, not a causal estimate of automation-displaced hours or an observed AI reinstatement ratio.

Supporting indicators. First, new-title and new-task emergence track work absent from a fixed classification baseline. Figure~\ref{fig:3} reports shares near 4 percent in 1990 for titles added during the 1980s, 11 percent in 2000 for titles added during the 1990s, and 10 percent in 2018 for titles added during 2000-2018. Second, employer formation and first hires locate the transition from entrepreneurial activity to paid work; applications, prototypes, and founder effort capture earlier stages. Third, intangible investment tracks software, data, training, and process redesign, the complements emphasized by Brynjolfsson, Rock, and Syverson (2021).

The uptake signature. Expansion can become visible in complementary investment, emerging activities and work content, and demand for human input at different dates. Figure~\ref{fig:13} reports AI-specific investment and role indicators. Following projects into sales, first hires, and paid hours would establish when additional activity becomes labor demand.

Measurement lags change what the ratio reads. If hours in additional activities enter the statistics only after a delay, as founder time, contractor work, and platform work become payroll at some rate, the measured series overstates displacement and understates reinstatement for years. Table~\ref{tab:A15} applies an exponential delay with a mean of one to three years to the new-activity hours of the simulations in Section 4.3. With the efficiency gain diffusing over seven years and \(\lambda\)=0.2, a true trough of 0.23 percent is measured at 0.6 to 1.0 percent and the measured recovery moves from 6.9 to between 7.7 and 11.7 years; with \(\lambda\)=0 the measured trough reaches 2 percent. The underlying paths are unchanged; only the record is late.

Two public series bound what payroll misses in the AI era. Census Nonemployer Statistics count businesses without employees: 30.4 million in 2023 with \$1.75 trillion in receipts, growing 2.7 percent a year over 2012-2023 and 2.1 percent in 2023. In 2024 the count rose 5.1 percent, or 1.56 million establishments, against a net private payroll gain of 0.89 million in the establishment flows of Figure~\ref{fig:12}. Valued at 2023 average receipts of \$57,600 and converted at 2023 average pay per payroll job of \$68,300, the 2024 increase is equivalent to about 1.3 million payroll jobs. Most nonemployer businesses are part-time, so the full-time equivalent is smaller, and the counts alone do not attribute the acceleration to AI; but on either conversion the non-payroll margin grew by more than the payroll margin in 2024 (U.S. Census Bureau, 2026). Second, human data work for AI training, the expert evaluation and annotation that AI laboratories purchase, is almost entirely contractor work outside payroll. Industry reporting puts contractor payments at the largest vendors in the low billions of dollars a year, with one vendor paying about 30,000 contractors more than \$4 million a day (New York Times, 2026); at average payroll compensation that is tens of thousands of full-time equivalents, new work concentrated in exposed occupations but small against payroll employment.

Measuring these stages together requires following project starts, prototypes, validation, first customers, and first hires, alongside task-level time savings. Failed and unfinished projects remain part of the record. The uptake measure follows where additional activity forms, when it becomes paid work, and how its scale compares with adjustment in established production.

\section{Implications}

The two-margin view calls for following established activities through task requirements, output, payroll, and vacancies, and new activity through development, entry, survival, and scale. Historical cohorts show why repeated entry matters even when particular cohorts contract. The model identifies how compute availability and unfinished complements shape that expansion.

Infrastructure and organization are part of the economic return to AI. Lower compute costs can make more task assignments and more projects profitable; reusable evaluation, integration, customer access, and financing can help feasible ideas become operating activity. Worker support and training address adjustment within existing work, while reducing barriers to viable new activity addresses the creation margin. The appropriate outcome set includes real wages, consumption, founder returns, paid hours, and worker counts rather than requiring every gain to appear as additional payroll.

\section{Conclusion}

AI changes both the cost of existing work and the range of activities worth developing. Firms automate when the full cost of an acceptable AI-assisted result is lower; entrepreneurial entry requires development and commercialization, with founder effort kept separate from paid work. Historical entry contributes sustained employment, but its accounting contribution alone does not identify a technology effect. In the simulations, new activity expands and real wages and consumption rise, while total hours depend on the balance of labor savings, entry, and household responses. Measuring these margins separately makes it possible to assess when expanded capabilities become demand for hired labor.

\clearpage
\clearpage
\clearpage
\begin{figure}[H]
\centering
\includegraphics[width=.96\textwidth]{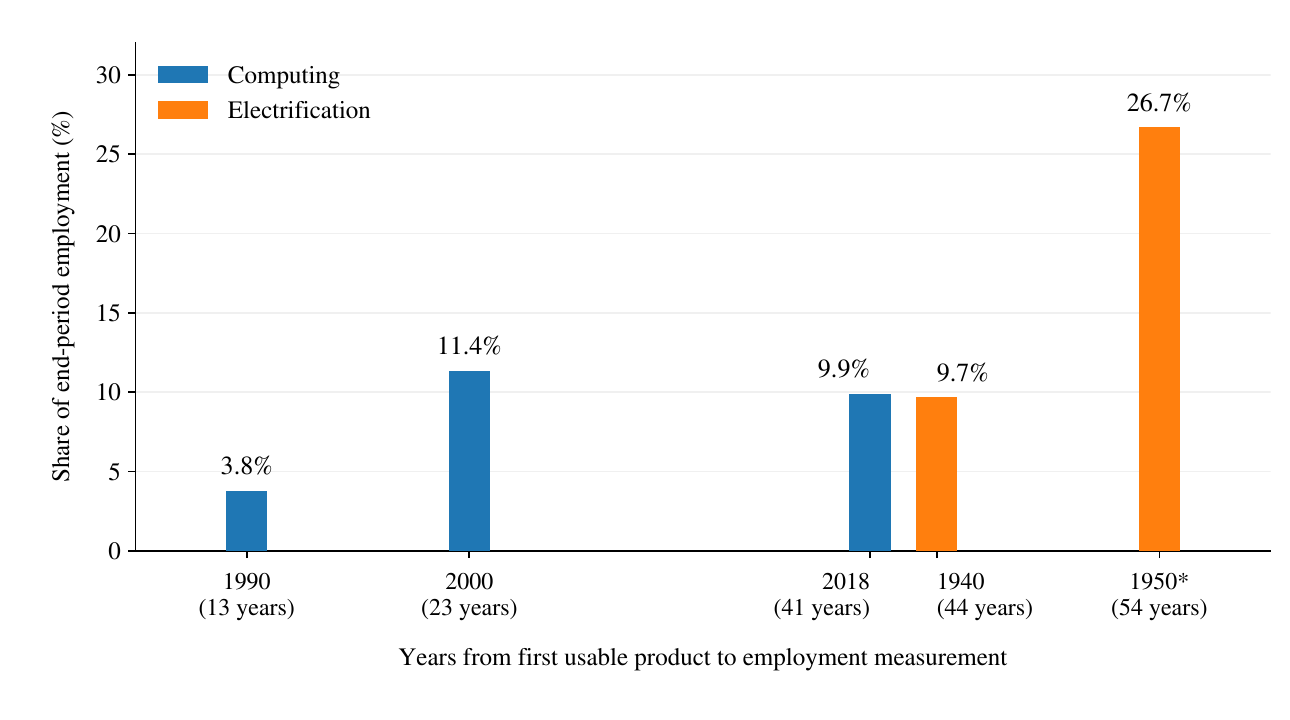}\\[5pt]
\caption{Employment in newly added job titles, by measurement year}\label{fig:3}
\begin{minipage}{.96\textwidth}\footnotesize Notes. Source: Autor, Chin, Salomons, and Seegmiller (2024) replication panel. Each bar is a period-specific share of end-period full-time-equivalent (FTE) employment allocated to newly recorded titles, weighted by occupation employment in the labeled year. The computing observations for 1990, 2000, and 2018 cover titles added during the 1980s, 1990s, and 2000-2018; the 1940 and 1950 electrification observations cover the 1930s and 1940s. For example, 11.4 percent is a share of 2000 employment, neither cumulative since the first usable product nor growth since the prior report. Bar positions and years in parentheses equal the measurement year minus 1977 for computing or 1896 for electrification. The 2018 interval spans 18 years; the 1950 index also reflects reorganization (*). Different employment denominators prevent summing the bars.\end{minipage}
\end{figure}
\clearpage
\begin{figure}[H]
\centering
\includegraphics[width=.96\textwidth]{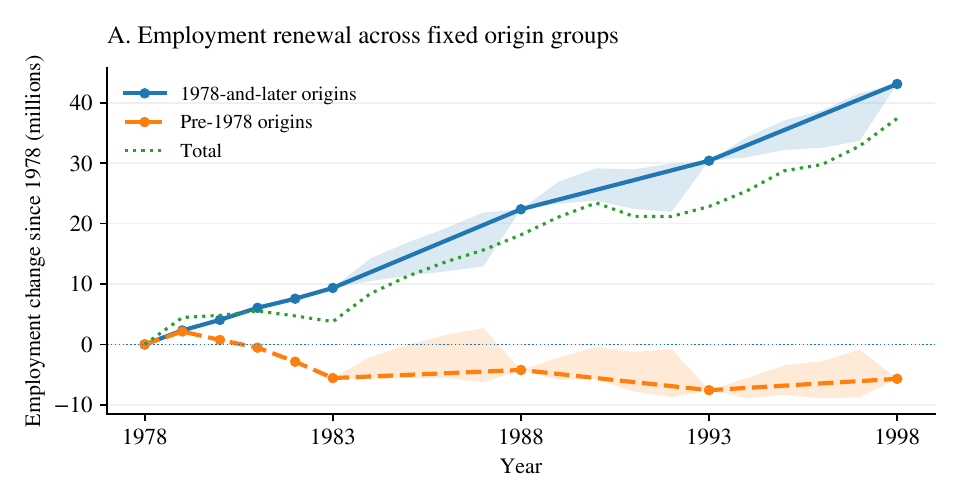}\\[5pt]
\includegraphics[width=.96\textwidth]{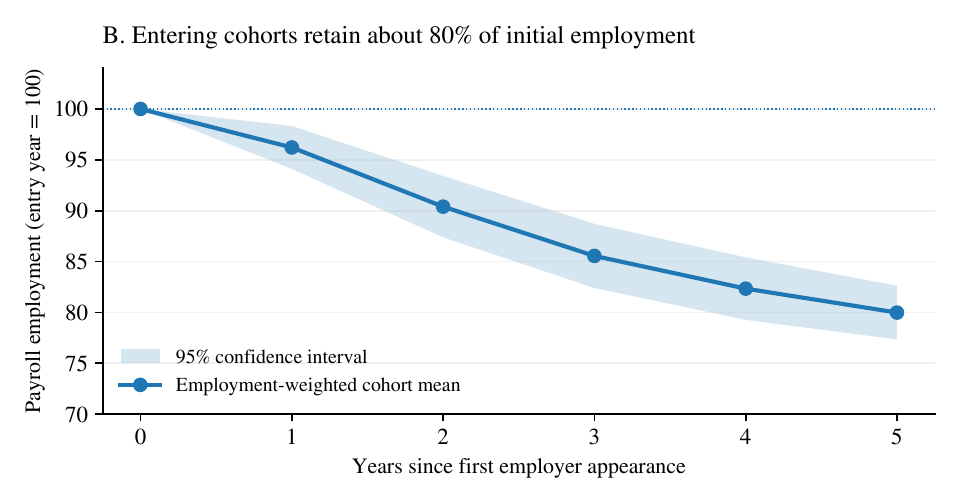}\\[5pt]
\caption{Employment accumulated by successive entrants and retained within cohorts}\label{fig:4}
\begin{minipage}{.96\textwidth}\footnotesize Notes. Source: Census BDS 2023 sector-by-firm-age tables. Panel A holds the origin boundary fixed at 1978 and subtracts each group's 1978 employment. Markers identify exact age-band-aligned dates, with lines connecting those anchors; shaded areas are allocation bounds when an age band crosses the cutoff. Panel B follows 1978-1993 entering cohorts through exact ages zero-five in 19 sectors, weighted by entry payroll. Its 95\% intervals use birth-cohort-clustered CR1 covariance and t(15) inference. Published stock accounting and regression uncertainty are different objects. Table~\ref{tab:A2} gives the fixed-origin totals; Table~\ref{tab:A4} reports cohort-retention estimates and robustness.\end{minipage}
\end{figure}
\clearpage
\begin{figure}[H]
\centering
\includegraphics[width=.96\textwidth]{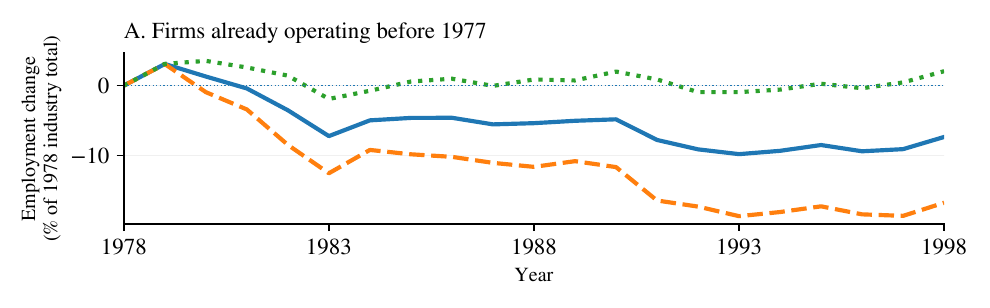}\\[5pt]
\includegraphics[width=.96\textwidth]{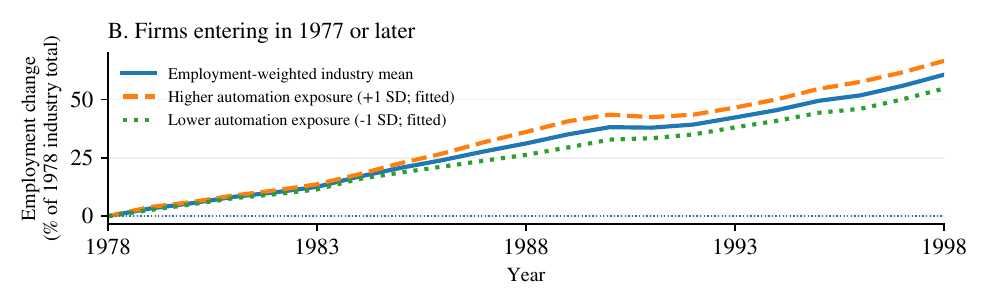}\\[5pt]
\includegraphics[width=.96\textwidth]{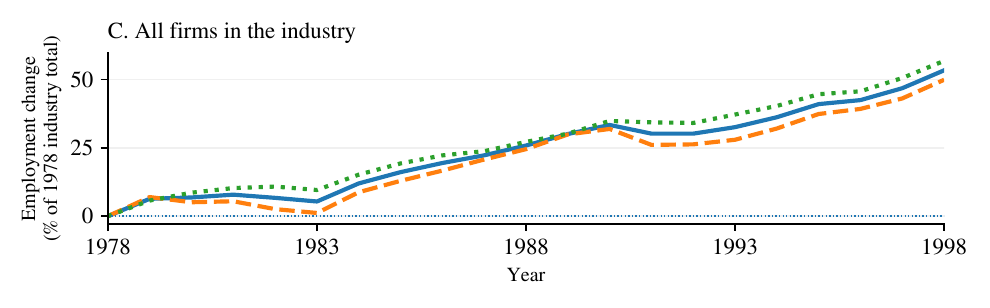}\\[5pt]
\caption{Incumbent and entrant employment by industry automation exposure, 1978-1998}\label{fig:5}
\begin{minipage}{.96\textwidth}\footnotesize Notes. Sources: Census BDS 2023 industry-age aggregates; IPUMS 1980 occupation-industry matrix; Autor et al. (2024) patent scores. Panels A and B follow firms first employing before 1977 and in 1977 or later, respectively; firms stay in the same group as they age. All panels show employment changes since 1978 as a percentage of total industry employment in 1978. The solid line is the mean across 84 industries, weighted by 1978 employment. The higher- and lower-exposure lines shift that mean by the fitted employment difference for exposure one standard deviation (SD) above or below its mean. The annual regressions hold initial industry size and two-digit sector effects fixed. These are fitted comparisons, not observed high/low-exposure group averages or confidence bounds. SD uses the full-sample 1978-employment weights; exposure combines 1980 occupations with 1980-1989 patent flows. Panels A and B add to Panel C for each line. Tables A5-A8 report the estimates and their standard errors.\end{minipage}
\end{figure}
\clearpage
\begin{figure}[H]
\centering
\includegraphics[width=.96\textwidth]{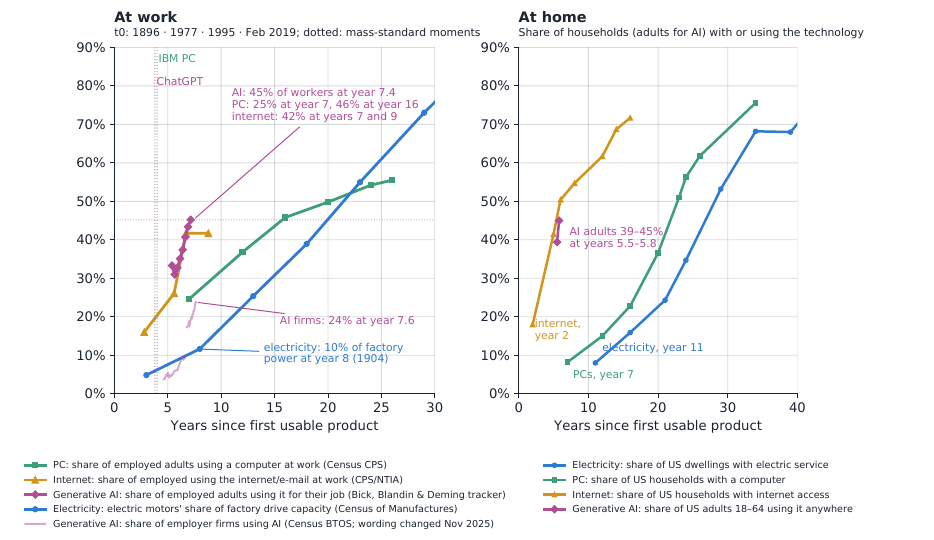}\\[5pt]
\caption{Four technologies on one clock}\label{fig:6}
\begin{minipage}{.96\textwidth}\footnotesize Notes. Each technology is aligned at the comparison's first usable product: Niagara-Buffalo transmission (1896), the first mass-market PCs (1977), the commercial internet (1995), and GPT-2 (February 2019). Dotted markers show the IBM PC (1981) and ChatGPT (November 2022). At work: Census CPS/BLS computer use; CPS/NTIA internet use; Bick, Blandin, and Deming's generative-AI tracker; manufacturing electric-motor capacity (Historical Statistics Dd848-853); and employer-business AI use (Census BTOS, with a November 2025 question change). At home: dwellings with electric service (Historical Statistics S108), households with computers or internet access (Census CPS), and adults 18-64 using generative AI. Denominators differ as stated in the legends.\end{minipage}
\end{figure}
\clearpage
\begin{figure}[H]
\centering
\includegraphics[width=.49\textwidth]{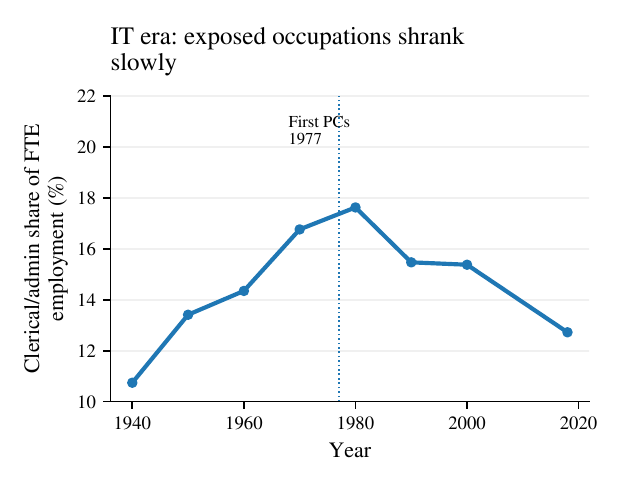}\hfill
\includegraphics[width=.49\textwidth]{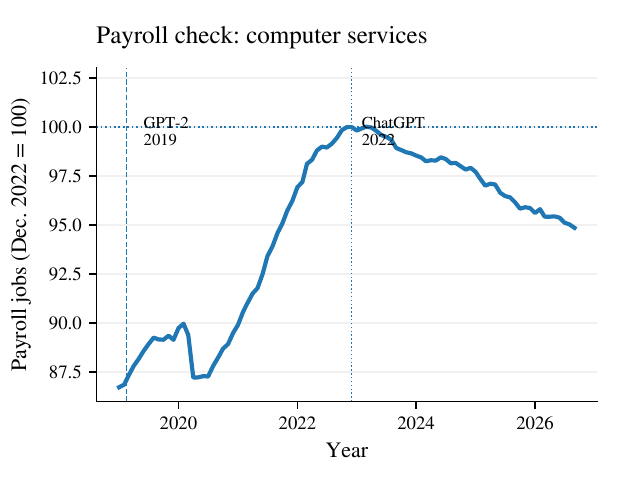}\\[7pt]
\caption{Displacement, now and then}\label{fig:7}
\begin{minipage}{.96\textwidth}\footnotesize Notes. Sources: Autor et al. (2024) replication panel; BLS CES6054150001, retained in the source workbook. Left: clerical and administrative share of total FTE employment, 1940-2018, with the first mass-market PCs marked in 1977. Right: seasonally adjusted paid payroll jobs in computer systems design and related services (NAICS 5415), December 2022=100, through September 2026; release markers indicate GPT-2 in February 2019 and ChatGPT in November 2022.\end{minipage}
\end{figure}
\clearpage
\begin{figure}[H]
\centering
\includegraphics[width=.96\textwidth]{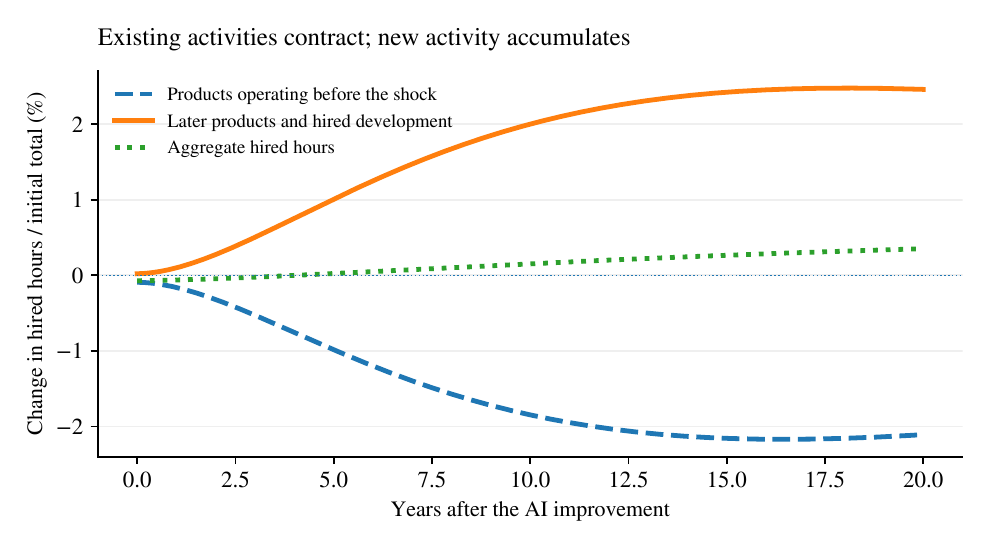}\\[5pt]
\includegraphics[width=.96\textwidth]{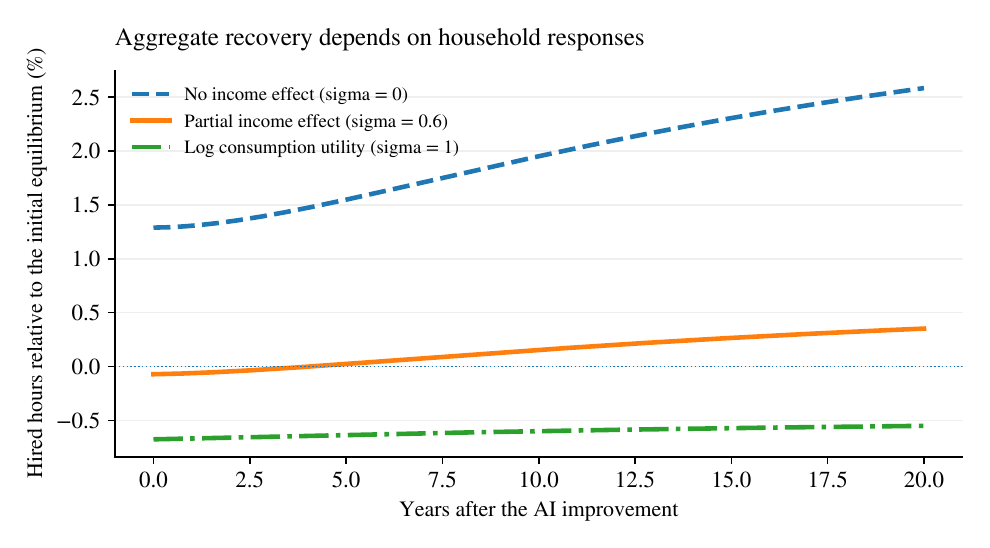}\\[5pt]
\caption{An AI shock through the existing-activity and creation margins}\label{fig:8}
\begin{minipage}{.96\textwidth}\footnotesize Notes. Model simulations of a permanent 50\% increase in compute efficiency and 25\% reduction in linear project-initiation effort. Top: partial income effects (sigma=0.6). The existing-product stock is M(0)exp(-delta t); its change in hired hours is measured relative to the same stock without the AI shock. The new-activity component includes later active products and hired development, net of their no-shock paths. The two contributions add exactly to the aggregate. Founder effort is excluded from hired hours. Bottom: identical technology shocks with sigma=0, 0.6, or 1. These are conditional preference scenarios, not confidence intervals. Table~\ref{tab:A10} includes sigma=0.5 and 0.75 as well.\end{minipage}
\end{figure}
\clearpage
\begin{figure}[H]
\centering
\includegraphics[width=.96\textwidth]{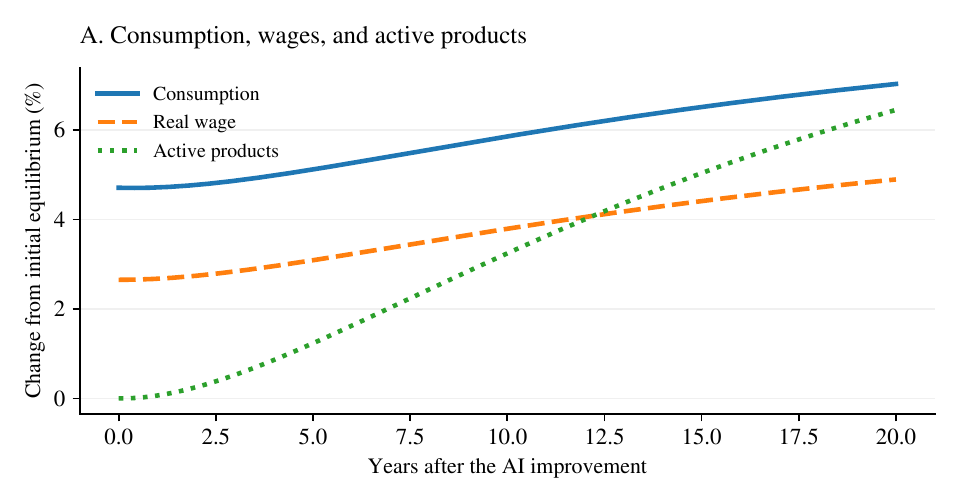}\\[5pt]
\includegraphics[width=.96\textwidth]{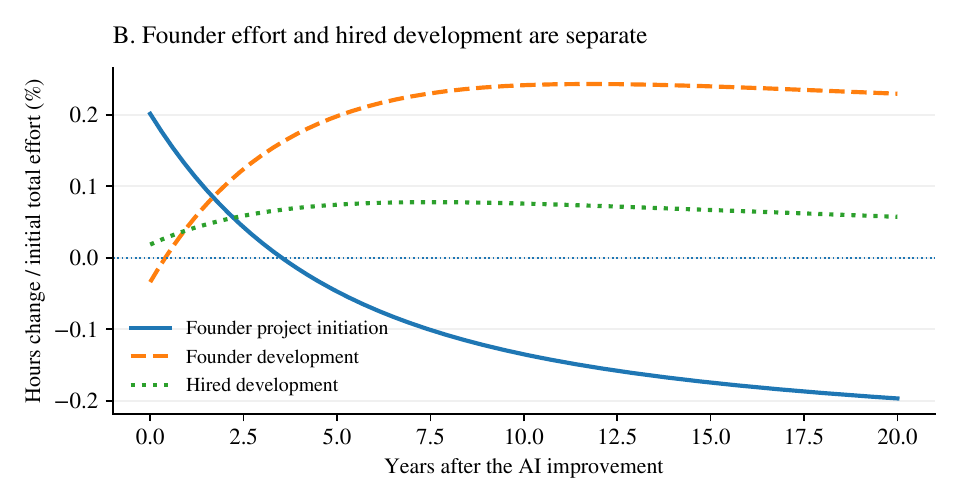}\\[5pt]
\caption{Living standards, entrepreneurial effort, and paid work}\label{fig:9}
\begin{minipage}{.96\textwidth}\footnotesize Notes. Model simulation under the sigma=0.6 partial-income-effect benchmark, matching Figure~\ref{fig:8} and Table~\ref{tab:A9}. Upper panel: consumption, real wages, and active product mass relative to the initial equilibrium. Lower panel: founder initiation, founder development, and hired development, each divided by initial combined effort. Table~\ref{tab:A10} reports alternative preferences and infrastructure. Founder effort is valued at its opportunity cost, not imputed as payroll.\end{minipage}
\end{figure}
\clearpage
\begin{figure}[H]
\centering
\includegraphics[width=.96\textwidth]{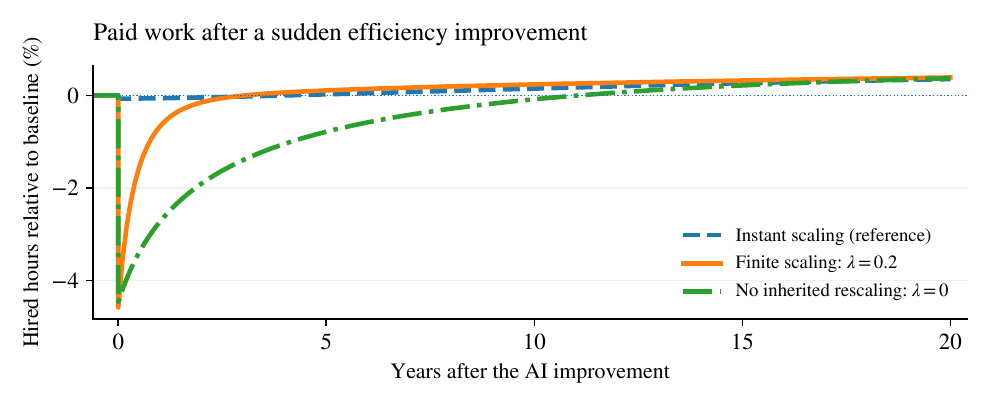}\\[5pt]
\includegraphics[width=.96\textwidth]{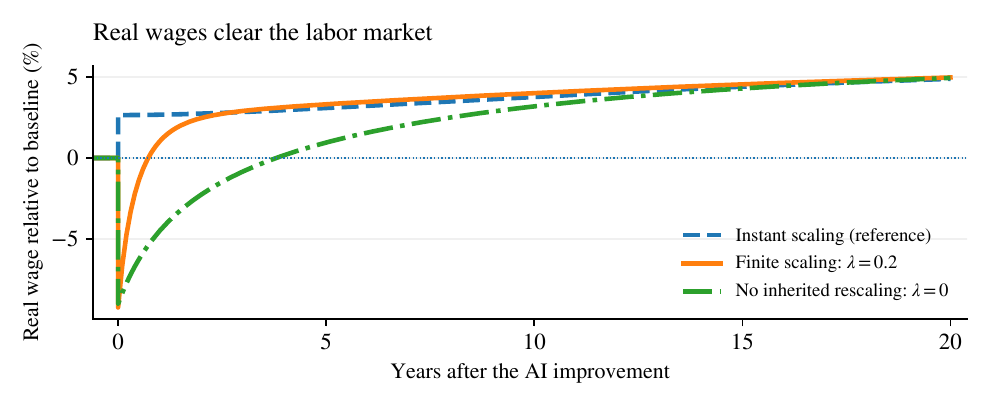}\\[5pt]
\caption{Incumbent scale adjustment after a sudden AI improvement}\label{fig:10}
\begin{minipage}{.96\textwidth}\footnotesize Notes. Model simulations with sigma=0.6. Effective compute efficiency rises 50 percent and linear initiation effort falls 25 percent at time zero. Finite \(\lambda\) constrains quantity per product for varieties operating before the improvement; later varieties choose scale flexibly. The infinite-speed case is the exact flexible-scale model. Both panels show percentage changes from the initial no-shock equilibrium; the vertical segment is the impact response. Founder effort is separate. Table~\ref{tab:A11} includes alternative adjustment speeds and sigma=1. The fixed-target half-life for \(\lambda\)=0.2 is 3.47 years, not an imposed recovery date.\end{minipage}
\end{figure}
\clearpage
\begin{figure}[H]
\centering
\includegraphics[width=.96\textwidth]{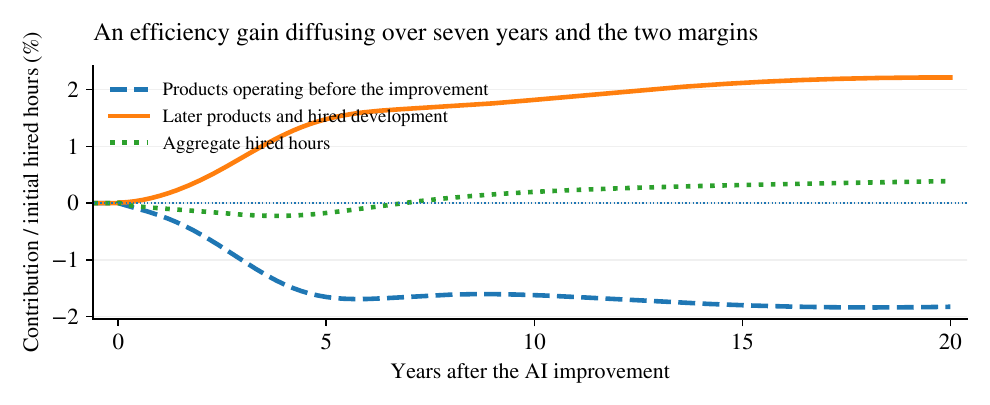}\\[5pt]
\includegraphics[width=.96\textwidth]{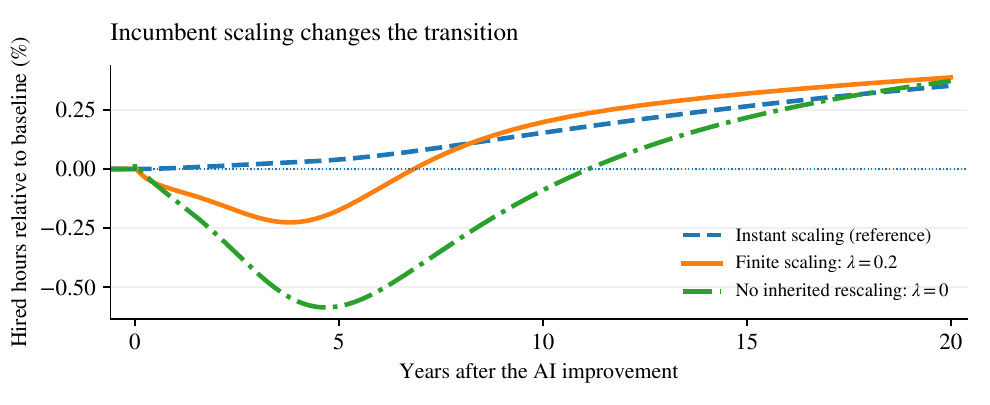}\\[5pt]
\caption{An efficiency gain diffusing over seven years and the two labor margins}\label{fig:11}
\begin{minipage}{.96\textwidth}\footnotesize Notes. Model simulations; sigma=0.6. Efficiency rises 50 percent along a normalized logistic path with location 3.5 years and scale 1.2 years: 50 percent completion at 3.62 years and 95 percent at 7.10 years. Linear initiation effort falls 25 percent at time zero. Adoption remains endogenous. Top: \(\lambda\)=0.2; actual inherited-product hours and later-product plus hired-development hours are measured against matched no-shock paths. The components sum to aggregate hired hours. Bottom: the same efficiency path with alternative adjustment speeds. Founder effort is separate; Tables A11-A12 report results.\end{minipage}
\end{figure}
\clearpage
\begin{figure}[H]
\centering
\includegraphics[width=.96\textwidth]{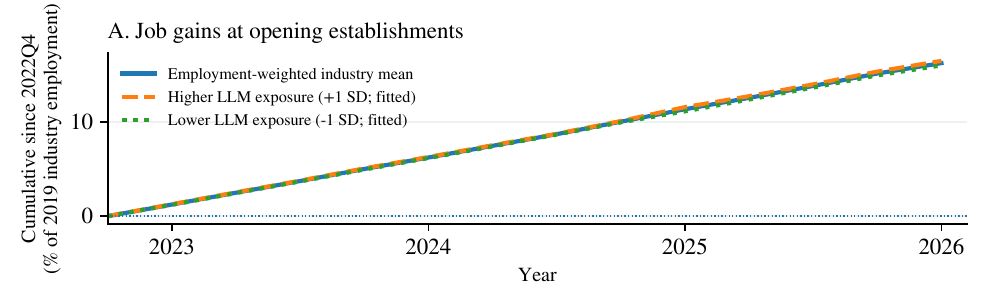}\\[5pt]
\includegraphics[width=.96\textwidth]{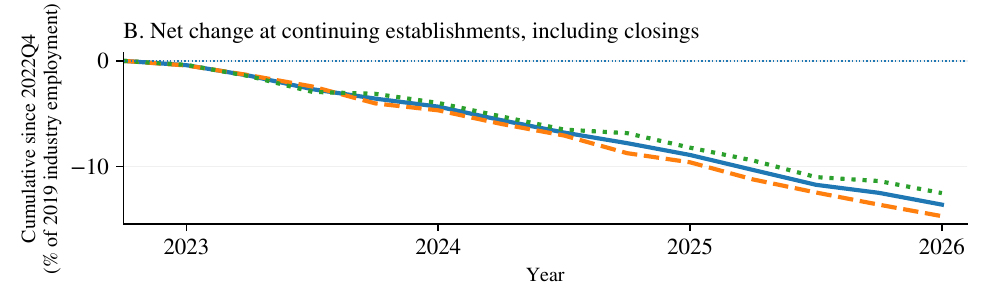}\\[5pt]
\includegraphics[width=.96\textwidth]{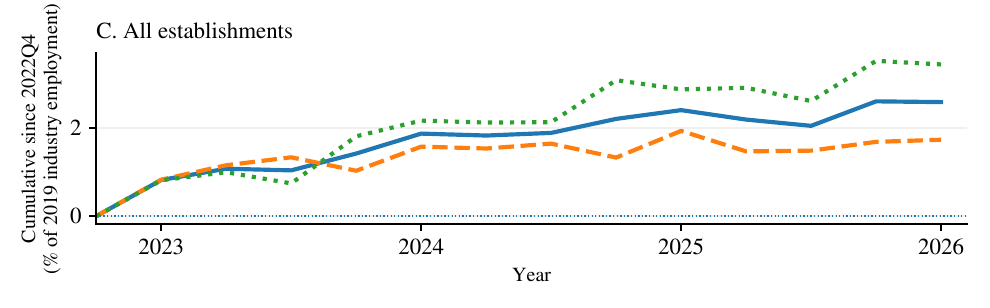}\\[5pt]
\caption{Establishment job flows by industry LLM exposure, 2022Q4-2026Q1}\label{fig:12}
\begin{minipage}{.96\textwidth}\footnotesize Notes. BLS Business Employment Dynamics, 72 three-digit NAICS industries with exposure, seasonally adjusted. Each panel shows the employment-weighted mean path of the outcome, cumulated from 2022Q4 as a percent of 2019 industry employment, and the fitted paths for industries one weighted standard deviation above and below the mean of LLM exposure, from quarter-by-quarter WLS with log 2019 employment and two-digit sector effects (Appendix F). Panel B includes closings. Table~\ref{tab:A14} reports the gradients.\end{minipage}
\end{figure}
\clearpage
\begin{figure}[H]
\centering
\includegraphics[width=.49\textwidth]{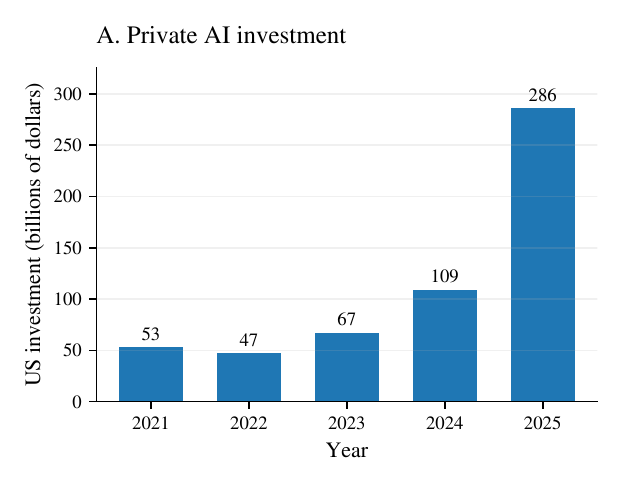}\hfill
\includegraphics[width=.49\textwidth]{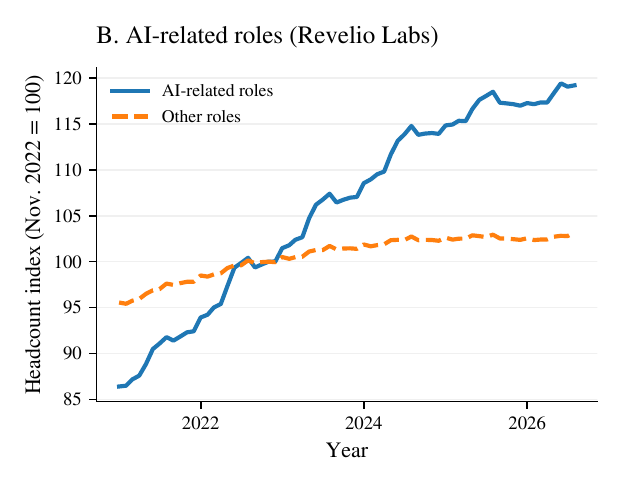}\\[7pt]
\caption{AI-era investment and emerging roles}\label{fig:13}
\begin{minipage}{.96\textwidth}\footnotesize Notes. Panel A: US private AI investment, Stanford AI Index / Quid; 2021-2023 values are rounded to the nearest billion as in the original series, and 2024-2025 are \$109.1 and \$285.9 billion. Panel B: Revelio Labs AI Labor Market Tracker, September 2026 release, headcount indices relative to November 2022 within its profile coverage.\end{minipage}
\end{figure}

\clearpage\phantomsection\section*{Appendix A. Model details, proofs, and numerical solution}\label{app:A}

Proof of Proposition 1. Compare the two costs of an accepted task result. Rearranging p\_X k/gamma + wv <= w gives the cutoff. Integration over adopted tasks gives the unit input requirements. At the cutoff both methods cost the same, so the envelope theorem removes a first-order boundary term from changes in unit cost. The numerical audit checks the cost inequality at the most compute-intensive adopted task.

Development-cost duality. Minimizing the total opportunity cost of founder and hired time subject to the normalized Cobb-Douglas input gives the unit cost and allocations in Section 3.4. Each input receives its technology share of total development cost. With founder-only development, effective effort equals founder effort and hired development is zero. This is the founder-only case in Table~\ref{tab:A10}. Founder effort is valued because it is scarce; no cash wage is imputed to it in the paid-hours accounting.

Proof of Proposition 2. For the new cohort, both extra stocks start at zero. The initial derivative of unfinished projects is Delta E. The active stock and its first derivative initially equal zero. Its second derivative initially equals nu times Delta E. Taylor expansion gives the stated orders. The result does not apply to an existing pipeline: higher effort can immediately change the completion flow from projects already under development. With founder-only development, the newly initiated cohort's paid operating input scales with the extra active stock, whereas a positive hired-development share adds input proportional to its unfinished stock.

Proof of Proposition 3. Subtract the initial hired-hours allocation from the current one. Split the change in operating hours into output-scale expansion valued at the current labor requirement, plus the change in the labor requirement valued at initial output. Rearranging gives the necessary and sufficient condition. The corresponding total-effort condition adds the change in founder effort to the left-hand side. Neither identity determines worker headcount without a separate hours-per-worker or employment margin.

Income effects and headcount. Differentiating hired-hours supply and the identity relating hours to paid positions gives:

\begin{equation}\label{eq:headcount}
d\ln H=\phi(d\ln w-\sigma d\ln C),\qquad N=H/\bar h,\qquad d\ln N=d\ln H-d\ln\bar h.
\end{equation}

A reduction in H can reflect fewer positions, fewer hours per position, or both. More consumption with less H is a feasible welfare improvement. The model does not choose an extensive employment margin, and the empirical BDS and CES job measures are not relabeled as model hours.

Numerical solution. Scale both asset values by marginal utility of consumption; denote the resulting values by \(\widehat{V}\) and \(\widehat{J}\). Their laws remove consumption growth from the discount term:

\begin{equation}\label{eq:scaled_values}
\begin{gathered}
\dot{\widehat{V}}=(\rho+\delta)\widehat{V}-C^{-\sigma}\pi, \\
\dot{\widehat{J}}=(\rho+\omega+\nu)\widehat{J}+C^{-\sigma}c_Dd-\nu\widehat{V}.
\end{gathered}
\end{equation}

Given (M,P,\(\widehat{V}\),\(\widehat{J}\)), a two-dimensional root solve determines p\_X/w and \(w_F\)/w. CES pricing determines w; compute supply and task input demand determine output and compute resources. The two household time conditions close the static allocation. Effort and initiation follow their optimality conditions. The four differential equations are solved as a boundary-value problem over 120 years, with initial stocks at their pre-shock steady state and terminal values at the post-shock steady state. Reported paths cover the first 25 years; main figures show 20. Table~\ref{tab:A9} supplies the parameters and the code retains unrounded values.

Calibration targets and sensitivity. The benchmark anchors M=1, AI task assignment=0.20, effective development effort=0.10, and \(w_F\)=w. Initiation curvature and the two time-disutility scales are recovered jointly with the compute scale. Changing theta redistributes initial development effort between founders and hired labor at equal initial opportunity costs. Changing eta preserves initial compute quantity, price, and production inputs but changes compute resource costs; percentage changes are always relative to the scenario's own initial consumption. Initial completion and abandonment imply an average time to the first project exit of about 4.8 years, with 2.7-7.7 years in the completion-speed comparisons. This duration is a scenario interpretation, not an estimated launch delay.

The replication audit checks markup pricing, both time-supply conditions, compute and goods clearing, both labor-allocation identities, and the development cost decomposition. The selected AI method never exceeds the human task cost at the cutoff. A no-shock experiment returns the original steady state. A longer-horizon, tighter-tolerance benchmark is reported in the audit file. Scenario variation is not assigned a probability distribution and is not described as a confidence band.

\clearpage\phantomsection\section*{Appendix B. Firm-age comparisons and cohort construction}\label{app:B}

Annual age-profile comparison. Table~\ref{tab:A3} uses sector-by-age-by-year observations from the Census BDS 2023 release. Define the net job-creation rate r as 100 times net job creation NJC divided by the published average-employment denominator D. The eight included categories are ages 0, 1, 2, 3, 4, 5, 6-10, and 11-15; ages 16+ form the omitted reference. We estimate:

\begin{equation}\label{eq:age_profile}
\begin{gathered}
r_{sgt}=100\frac{\mathrm{NJC}_{sgt}}{D_{sgt}}, \\
r_{sgt}=\beta_0+\sum_{a\in\mathcal{A}}\beta_a\mathbf{1}\{g=a\}+\vartheta^{\prime}Z_t+\lambda_s+u_{sgt}.
\end{gathered}
\end{equation}

Here A denotes the eight included age categories, Z contains annual nominal-GDP and nonfarm-business productivity log growth, and \(\lambda\) denotes sector effects. The four columns include neither adjustment, macro controls only, sector effects only, and both. The 1992-1998 sample has 1,197 observations: 19 sectors, nine age groups including the reference, and seven years. From 1992 all pre-1977 left-censored firms are at least sixteen years old. We pool job flows and denominators across the older age bands before computing the reference rate. Structurally unavailable bands contribute zero; unexplained missing values stop the calculation.

Weights equal each cell's published denominator. Standard errors cluster by sector with CR1 correction and t(18) inference. Table~\ref{tab:A3} estimates differences in annual rates across contemporaneous age groups, not employment retained by fixed cohorts. The approximately 200 percent entrant rate follows from zero previous employment under the symmetric growth denominator. Birth-cohort effects are not included because the broad age categories combine several cohorts.

Fixed origin. Census firm age is tied to employer history and the oldest establishment at firm birth. Firms present before the historical start are left-censored. The principal cutoff classifies origins before 1978 separately from 1978 onward. For each age band in year t, all employment belongs to the later group if its upper endpoint is at most t minus 1978, and to the earlier group if its lower endpoint exceeds that cutoff. Otherwise the entire cell is potentially on either side. The resulting lower and upper allocations exhaust the observed total jointly:

\begin{equation}\label{eq:bounds}
N_t^{\mathrm{lo}}=\sum_{a_{hi}\leq t-1978}\mathrm{Emp}_{a,t},\qquad N_t^{\mathrm{hi}}=N_t^{\mathrm{lo}}+\sum_{a_{lo}\leq t-1978<a_{hi}}\mathrm{Emp}_{a,t}.
\end{equation}

Earlier-origin bounds are total employment minus the opposite later-origin bound. Changes subtract the exact 1978 stocks. Single-year ages through five and five-year bands thereafter make the cutoff exact at the dates in Table~\ref{tab:A2}. Shading in Figure 4A is the nonparametric allocation range for the crossing age band, not a statistical confidence interval. Structural X cells correspond to impossible ages since 1977 and contribute zero; the code stops on unexpected nonstructural missing employment. Source rows, age mappings, and every annual bound are retained.

Scope of the cohort measure. This is a fixed employer-origin classification using public Census age lineages. Mergers, acquisitions, and sector changes can reassign establishments across published firm-age or industry cells, so the aggregate series is not an ownership-invariant matched-firm survival panel. Sector totals are summed for employment, not for national firm counts. National full-age tables collapsed across the two patent statuses independently reproduce all exact stock anchors. A pre-1977 left-censored versus 1977-and-later definition provides an annual robustness series without split age bands.

Cohort-retention design. For births 1978-1993, exact ages zero through five are available in each sector. The age-five outcome retains the cohort as it ages and includes contraction and closure, rather than conditioning on surviving establishments. The regression uses a common age profile, sector and birth-cohort effects, and annual macro controls; saturated calendar-year effects are not added because age, period, and cohort are mechanically linked. CR1 cluster covariance is calculated by birth cohort, and inference uses 15 degrees of freedom. Alternative sector clustering uses 18. Confidence intervals in Figure 4B describe the regression's variation across entering cohorts, not Census sampling errors around the national employment totals.

Table~\ref{tab:A4} retains the separate cohort-retention outcome: 100 times sector-cohort employment at age five divided by its own age-zero employment, reported as the difference from age zero. The baseline rows vary macroeconomic controls and sector and birth-cohort effects; the remaining rows vary weights, clustering, and sample windows. These estimates use 1978-1993 birth cohorts through 1998 and entry-employment weights, rather than the annual age-profile sample in Table~\ref{tab:A3}.

\clearpage\phantomsection\section*{Appendix C. Historical new-work accounting}\label{app:C}

The historical comparison uses 306 consistently coded occupations in the 1990-2018 portion of the Autor et al. (2024) replication panel. For each occupation, the share of 2018 titles added in the 2000 and 2018 Census Alphabetical Index volumes is capped at one and multiplied by its 2018 full-time-equivalent (FTE) labor input. Summing gives 32.37 million imputed new-work FTE units, 20.14 percent of 160.73 million end-period FTE units.

The absolute-loss denominator sums reductions in FTE employment among occupations that shrink between 1990 and 2018. It equals 11.09 million, giving a ratio of 2.92. The relative-loss denominator compares each occupation's 2018 FTE employment with its 1990 FTE employment multiplied by aggregate FTE growth. Summing negative deviations gives 29.60 million and a ratio of 1.09. Neither denominator isolates hours displaced by automation. The comparison measures observed gains and losses using the same underlying panel and fixed occupational definitions.

Figure~\ref{fig:3} uses the employment-weighted new-title share in each Census index volume and labels bars by the employment measurement year. Shares are 3.79 percent in 1990 (titles added during the 1980s), 11.36 percent in 2000 (1990s), and 9.91 percent in 2018 (2000-2018). The 1940 and 1950 observations are 9.68 and 26.70 percent (titles added during the 1930s and 1940s). Horizontal coordinates are measurement year minus the first-usable-product year: 13, 23, and 41 years after 1977 for computing; 44 and 54 years after 1896 for electrification. The bars retain the same period-specific employment allocation and are not cumulative or employment growth rates. Classification changes and unequal lookback intervals are recorded in the figure note. Only historical employment-based bars are plotted; current AI-role counts appear separately in Figure~\ref{fig:13}.

\clearpage\phantomsection\section*{Appendix D. Incumbent scale adjustment and verification}\label{app:D}

Scope. The extension implements the supplied quantity-adjustment law for products operating when AI improves. Later products choose their scale flexibly. The parameter \(\lambda\) is a reduced-form speed; no adjustment-cost expenditure or optimal investment condition is added. Household optimality, accepted-output costs, development, project laws, and compute resources remain unchanged. The symmetric pricing condition applies only to the flexible-scale reference.

Values and accounting. The state for inherited output affects consumption and equilibrium prices, while the value of a newly completed product drives development and initiation. The profits of pre-shock products accrue to the household; they need not enter the new-project recursion to matter for household resources. At finite \(\lambda\), initial physical output and the active and unfinished stocks are predetermined. Development effort can jump, so:

\begin{equation}\label{eq:scale_impact}
H_{0+}-H_0=Q_0(\ell_{0+}-\ell_0)+P_0(h_{0+}-h_0).
\end{equation}

In the sudden-improvement case with \(\lambda\)=0.2, the first term is -4.75 percent of initial hired hours and the second is +0.17 percentage points, giving -4.58 percent. At later dates, inherited-product employment separates into task savings and output rescaling:

\begin{equation}\label{eq:scale_actual_decomp}
\begin{gathered}
\Delta H_t^I=M_t^I[(\ell_t-\ell_0)q_0+\ell_t(q_t^I-q_0)], \\
\Delta H_t^N=\Delta H_t-\Delta H_t^I.
\end{gathered}
\end{equation}

Both components are compared with the corresponding no-shock paths. The second includes later-product operating and maintenance hours plus hired development, not founder effort. The decomposition adds exactly to the aggregate paid-hours change.

Optional efficiency diffusion. Let G(t) be a logistic time profile, distinct from the distribution of task requirements. Normalize the efficiency path to start at its initial value:

\begin{equation}\label{eq:scale_logistic}
\begin{gathered}
G(t)=\frac{1}{1+\exp[-(t-3.5)/1.2]}, \\
\gamma_t=1+0.5\frac{G(t)-G(0)}{1-G(0)}.
\end{gathered}
\end{equation}

The normalization gives a half-completion date of 3.62 years rather than the unnormalized logistic location of 3.5. Task assignment is still selected from relative costs at each date. The reduction in linear initiation effort is available immediately, including when the efficiency gain diffuses over seven years.

Solution. The finite-speed boundary-value problem has five states: active products, unfinished projects, marginal-utility-scaled operating and project values, and inherited per-product quantity. The static block solves for the compute-to-wage ratio, founder-to-hired-wage ratio, and later-product quantity. Inherited quantity and the two stocks begin at the pre-shock steady state; terminal scaled values equal their post-shock values. We use a 120-year horizon and collocation tolerance 3e-5, and report years 0-25. A separate 180-year, 3e-6 check assesses the terminal approximation for the two featured \(\lambda\)=0.2 cases.

Limit cases and audits. The \(\lambda\)=0 case is solved exactly. Infinite speed uses the original four-state symmetric system, allowing time-varying efficiency where applicable. A large finite speed retains predetermined quantity at the instant of the shock and is not a valid substitute for the infinite-speed impact. The code checks the CES quantity and revenue identities, new-product markup, compute and goods clearing, household conditions, and both labor decompositions. Recovery is the first upward zero crossing after a negative excursion, interpolated on the 0.1-year reporting grid; the code does not assert crossings outside the reported horizon.

\clearpage\phantomsection\section*{Appendix E. Industry exposure, accounting, and inference}\label{app:E}

Observation unit and inputs. The analysis uses 84 three-digit industries in 19 two-digit sectors for 1978-1998 (1,764 industry-years). The supplied panel excludes central banking (521) and pipeline transportation (486) because of suppressed employment cells. Its pre-1977 group is the BDS left-censored category; the complement contains 1977-and-later employer origins. These are firm-age lineages in published industry aggregates, not ownership-invariant matched firms. The replication checks each group against the included industry totals and retains a single balanced sample.

Exposure construction. The supplied IPUMS matrix contains 50,714 occupation-industry cells and 98.99 million weighted employed people aged 16 or older. Occupation scores cover approximately 98.3 percent of that employment. The automation measure is the employment-weighted mean of log(1+automation-patent exposure), using the 1990 row of the later-period Autor et al. (2024) occupational panel and therefore patent flows dated 1980-1989. The DOT STS measure is routine-cognitive intensity; finger dexterity captures routine-manual content (Autor, Levy, and Murnane, 2003; Autor and Dorn, 2013). Neither is literally a percentage of tasks. The supplied occupation scores, matrix, and industry crosswalk reproduce the included industry exposure file to numerical precision.

\begin{equation}\label{eq:industry_exposure}
R_i=\sum_o s_{oi,1980}r_o,\qquad z_i=\frac{R_i-\overline{R}_w}{s_{R,w}}.
\end{equation}

The crosswalk chains Census 1990, 2000, 2012, and 2017 industries into 2017 NAICS. Its original construction uses first listed concordances for some many-to-many steps, equal splits across named NAICS groups, and 1978 BDS employment for broad partial-sector categories. These are allocation assumptions, not observed historical four-digit employment. The supplied chain is preserved and audited rather than represented as an independently validated raw-data crosswalk. The aggregated IPUMS matrix includes public and self-employed workers, while BDS records covered employer jobs; class-of-worker-specific composition cannot be recovered from this matrix. Both 1980 composition and 1980s patents overlap the outcome window.

Comparable employment accounting. Write I for employment in pre-1977 origins, N for 1977-and-later origins, and T=I+N. Every outcome uses the same initial total employment denominator and subtracts the initial stock, including the later-origin employment already present in 1978:

\begin{equation}\label{eq:industry_accounting}
\begin{gathered}
D_{it}^I=100\frac{I_{it}-I_{i,1978}}{T_{i,1978}},\qquad D_{it}^N=100\frac{N_{it}-N_{i,1978}}{T_{i,1978}}, \\
D_{it}^T=D_{it}^I+D_{it}^N.
\end{gathered}
\end{equation}

These common-denominator outcomes allow valid addition of the two margins. The supplied alternative incumbent and total outcomes are log changes; they are reproduced in the results archive but are not divided into the later-origin contribution to calculate an offset percentage. In the included sample, the employment-weighted 1998 contributions are -7.33 percent for incumbents and +60.72 percent for later origins, totaling +53.39 percent. The weighted mean incumbent log change is -15.95 log points and the mean total log change is +33.54; those use a different transformation and are not arithmetic percentage changes.

\begin{equation}\label{eq:industry_reg}
D_{it}^m=\alpha_t+\beta_t^m z_i+\gamma_t^m\ln T_{i,1978}+\alpha_{s(i),t}^m+u_{it}^m.
\end{equation}

Regressions are estimated separately by year using fixed 1978 employment weights. Tables A5-A6 add initial size and sector fixed effects separately and jointly; common annual macro variables are absorbed by each cross-section intercept. Each regression has 84 observations. Full specifications have 63 residual degrees of freedom. Because the outcomes, observations, weights, and design matrix coincide, the three estimated exposure slopes satisfy the same adding-up identity as employment. In Figure~\ref{fig:5}, the solid line is the 1978-employment-weighted outcome mean. The higher-exposure line adds the annual fitted slope for one standard deviation of exposure, and the lower-exposure line subtracts it. Thus, the fitted lines differ only in exposure; sector effects and initial employment are held fixed. They are comparisons anchored to the weighted mean, not observed high/low group averages or confidence bounds.

Inference and robustness. Table~\ref{tab:A7} reports annual gradients with pointwise HC1 standard errors; these are not simultaneous bands or a test of relative onset dates. Table~\ref{tab:A8} reports sector-clustered CR1 inference with t(G-1), equal weighting, exclusion of manufacturing, joint augmentation exposure, alternative task measures, HC3 after removing singleton sectors, and changes beginning in 1985. The reproduction also saves leave-one-sector-out estimates. Few sector clusters and overlap of patent timing with outcomes remain features of the design; HC1 does not eliminate within-sector correlation.

Interpretation. Automation-linked industries show an early negative incumbent gradient, while later-origin employment has positive but imprecise gradients. No individual annual later-origin gradient reaches the 10 percent level under the full HC1 specification. This does not establish a three-to-five-year causal delay. By 1998, dividing the common-unit point estimates gives an offset of about 0.63, but the numerator and denominator are imprecise (delta-method standard error about 0.89). Accordingly, the paper reports the margins rather than treating that ratio as an estimated reinstatement fraction. Sector reassignment, general business cycles, and other industry shocks can contribute to the associations.

Replication boundary. The package rebuilds the regression analysis, exposure aggregation, figures, and tables from the supplied aggregated inputs. The raw IPUMS microdata, source concordance workbooks, and detailed BDS source file were not included in the uploaded test archive. The original source program and memorandum are preserved separately. The audit records exact reproduction of the supplied slopes, input coverage, adding-up checks, and corrections to units, cutoff wording, and inference.

\clearpage\phantomsection\section*{Appendix F. Establishment job flows by LLM exposure, 2019-2026}\label{app:F}

This appendix applies the industry design of Section 2.5 and \hyperref[app:E]{Appendix E} to the AI era, using establishment job flows in place of firm-age payroll. Outcomes come from the BLS Business Employment Dynamics, three-digit NAICS, seasonally adjusted, 1992Q3-2026Q1 (October 28, 2026 release), which report job gains at opening establishments, expansions at continuing establishments, contractions, and closings. Employment levels are recovered from the published rates, whose denominator is the average of current and previous-quarter employment. Three outcomes are cumulated from 2022Q4, the quarter of ChatGPT's release, and expressed as a percent of 2019 average employment: job gains at openings, the entry margin; expansions minus contractions minus closings, the adjustment at continuing establishments including exits; and their sum. Gradients computed from a 2019Q4 base, which include the pandemic, are reported in the replication files and are near zero as well. Of 75 industries, 72 have exposure.

Exposure is the employment-weighted mean of Eloundou et al. (2024) GPT-4 beta exposure over each industry's May 2022 occupational composition from the BLS Occupational Employment and Wage Statistics, aggregated from four-digit to three-digit NAICS; 92.7 percent of OEWS employment carries a score. Beta is the share of an occupation's tasks for which a large language model, or software built on one, could halve completion time. Standardized with 2019 BED employment weights, the measure has mean 0.34 and standard deviation 0.125. It describes the tasks an industry's employees perform, not the tasks involved in founding a firm. Marchesi and Tang (2026) publish only the top and bottom 20 of their 81 founding-task scores; 27 of those 40 industries are in the BED panel, and Table~\ref{tab:A14} uses them as an indicator without sector effects. Across the 36 industries where both measures exist they correlate at 0.77.

The specification is the quarterly analogue of equation (31): for each quarter q and outcome m, the cumulative flow since 2022Q4 is regressed across industries on standardized exposure, log 2019 employment, and two-digit sector effects, with 2019 employment as weights and HC1 standard errors, so that the exposure slope is directly comparable with Tables A5-A8. Figure~\ref{fig:12} adds each quarter's gradient to the employment-weighted mean path from 2022Q4, as Figure~\ref{fig:5} does from 1978. The gradients are near zero at every quarter on both margins. Cumulative from 2022Q4 to 2026Q1, one standard deviation of exposure is associated with 0.24 percent more job gains at openings (standard error 1.91), 1.09 percent less net employment at continuing establishments (1.85), and 0.85 percent less in total (0.98); the typical industry gained 16.2 percent at openings and lost 13.6 percent at continuing establishments over the same period. Gradients over 2019Q4-2022Q4 are also near zero, so no pre-trend is masked. In the Marchesi-Tang comparison the compatible industries show more job gains at openings since 2022Q4, 5.69 percent of 2019 employment with a standard error of 5.72, and weaker continuing-establishment employment; without sector effects the latter reflects sector trends in retail, wholesale, and information relative to health care and construction and is not read as an AI effect. The full founding-task index would allow the within-sector test; it has been requested from the authors.

\clearpage
\setcounter{figure}{0}\renewcommand{\thefigure}{A\arabic{figure}}
\setcounter{table}{0}\renewcommand{\thetable}{A\arabic{table}}
\begin{table}[H]
\centering
\caption{Annual net job flows by contemporaneous firm age, 1982-1998}\label{tab:A1}
\small
\begin{tabular}{>{\raggedright\arraybackslash}p{0.1350\textwidth}>{\centering\arraybackslash}p{0.1530\textwidth}>{\centering\arraybackslash}p{0.1530\textwidth}>{\centering\arraybackslash}p{0.1530\textwidth}>{\centering\arraybackslash}p{0.1530\textwidth}>{\centering\arraybackslash}p{0.1530\textwidth}}
\toprule
Year & Entrants (0) & Young (1-5) & Older (6+) & National total & Residual \\
\midrule
1982 & 2.272 & -0.803 & -2.156 & -0.705 & -0.018 \\
1983 & 2.518 & -0.720 & -2.711 & -0.935 & -0.023 \\
1984 & 2.799 & 0.197 & 1.822 & 4.796 & -0.022 \\
1985 & 2.941 & -0.204 & 0.159 & 2.879 & -0.016 \\
1986 & 2.978 & -0.496 & 0.014 & 2.479 & -0.017 \\
1987 & 3.402 & -0.501 & -1.202 & 1.679 & -0.020 \\
1988 & 3.055 & -0.610 & -0.044 & 2.384 & -0.018 \\
1989 & 2.999 & -0.325 & 0.244 & 2.902 & -0.016 \\
1990 & 2.886 & -0.549 & -0.161 & 2.163 & -0.014 \\
1991 & 2.540 & -1.427 & -3.349 & -2.250 & -0.015 \\
1992 & 2.491 & -0.631 & -1.812 & 0.033 & -0.016 \\
1993 & 2.539 & -0.246 & -0.656 & 1.622 & -0.016 \\
1994 & 2.691 & -0.384 & -0.073 & 2.216 & -0.018 \\
1995 & 2.790 & -0.102 & 0.807 & 3.477 & -0.018 \\
1996 & 2.796 & -0.383 & -0.591 & 1.804 & -0.018 \\
1997 & 2.808 & -0.239 & 0.184 & 2.736 & -0.017 \\
1998 & 2.782 & -0.104 & 0.682 & 3.347 & -0.014 \\
Annual mean & 2.782 & -0.443 & -0.520 & 1.802 & -0.017 \\
\bottomrule
\end{tabular}
\vspace{6pt}
\begin{minipage}{.96\textwidth}\footnotesize Notes. Millions of payroll jobs per year. Source: Census BDS 2023 release; firm-age totals aggregate the two mutually exclusive patent statuses in bds2023\_pfg\_utly\_fa.csv. National totals use bds2023.csv. Residual equals the independently published national total minus the three age-group sums. Membership changes as firms age; these annual flow groups are distinct from the fixed-origin stocks in Table~\ref{tab:A2}. Entrant flows are positive at first employer appearance; the comparison shows their magnitude relative to incumbent net flows.\end{minipage}
\end{table}

\clearpage
\begin{table}[H]
\centering
\caption{Employment in fixed groups of firm-entry years}\label{tab:A2}
\small
\begin{tabular}{>{\raggedright\arraybackslash}p{0.0900\textwidth}>{\centering\arraybackslash}p{0.1620\textwidth}>{\centering\arraybackslash}p{0.1620\textwidth}>{\centering\arraybackslash}p{0.1620\textwidth}>{\centering\arraybackslash}p{0.1620\textwidth}>{\centering\arraybackslash}p{0.1620\textwidth}}
\toprule
Year & 1978+ stock & Pre-1978 stock & Change: 1978+ & Change: pre-1978 & Total change \\
\midrule
1978 & 2.551 & 67.704 & 0.000 & 0.000 & 0.000 \\
1979 & 4.870 & 69.844 & 2.319 & 2.140 & 4.459 \\
1980 & 6.621 & 68.445 & 4.070 & 0.741 & 4.811 \\
1981 & 8.617 & 67.171 & 6.066 & -0.533 & 5.534 \\
1982 & 10.125 & 64.849 & 7.574 & -2.855 & 4.719 \\
1983 & 11.906 & 62.135 & 9.355 & -5.569 & 3.786 \\
1988 & 24.928 & 63.474 & 22.377 & -4.230 & 18.147 \\
1993 & 32.974 & 60.129 & 30.423 & -7.576 & 22.847 \\
1998 & 45.678 & 62.011 & 43.127 & -5.693 & 37.434 \\
\bottomrule
\end{tabular}
\vspace{6pt}
\begin{minipage}{.96\textwidth}\footnotesize Notes. Millions of payroll jobs. Source: Census BDS 2023 sector-by-firm-age tables, summed across 19 sectors. Firm-origin groups are defined once: origin year (calendar year minus reported age) of 1978 or later versus before 1978, using Census firm-age lineage. Changes subtract each group's 1978 employment, including 2.551 million jobs already at 1978 entrants. Dates shown are exactly identified by the published age bins. National full-age BDS-PF totals, summed across the mutually exclusive patent statuses, reproduce these stocks exactly. Annual bounds are supplied in the replication files. These are published-stock accounting totals; no sampling confidence bands are assigned.\end{minipage}
\end{table}

\clearpage
\begin{table}[H]
\centering
\caption{Annual net job-creation rates by firm age}\label{tab:A3}
\small
\begin{tabular}{>{\raggedright\arraybackslash}p{0.3150\textwidth}>{\centering\arraybackslash}p{0.1463\textwidth}>{\centering\arraybackslash}p{0.1463\textwidth}>{\centering\arraybackslash}p{0.1463\textwidth}>{\centering\arraybackslash}p{0.1463\textwidth}}
\toprule
 & (1) Raw & (2) Macro & (3) Sector FE & (4) Both \\
\midrule
Age 0 & 200.06*** & 200.05*** & 200.23*** & 200.22*** \\
 & (0.35) & (0.35) & (0.61) & (0.61) \\
Age 1 & -0.12 & -0.12 & 0.00 & -0.01 \\
 & (3.06) & (3.06) & (2.80) & (2.80) \\
Age 2 & -3.87*** & -3.86*** & -3.80*** & -3.79*** \\
 & (1.26) & (1.26) & (1.04) & (1.04) \\
Age 3 & -3.17*** & -3.16*** & -3.12*** & -3.11*** \\
 & (0.81) & (0.80) & (0.72) & (0.72) \\
Age 4 & -2.74*** & -2.73*** & -2.71*** & -2.71*** \\
 & (0.88) & (0.89) & (0.87) & (0.87) \\
Age 5 & -2.03** & -2.01** & -2.03** & -2.02** \\
 & (0.76) & (0.76) & (0.80) & (0.81) \\
Age 6-10 & -1.16* & -1.18* & -1.19* & -1.21* \\
 & (0.65) & (0.65) & (0.58) & (0.58) \\
Age 11-15 & -0.35 & -0.33 & -0.37 & -0.35 \\
 & (0.49) & (0.49) & (0.44) & (0.44) \\
Nominal GDP growth & - & -0.42* & - & -0.42* \\
 &  & (0.22) &  & (0.22) \\
Labor productivity growth & - & -0.25** & - & -0.25** \\
 &  & (0.11) &  & (0.11) \\
Omitted age group & 16+ & 16+ & 16+ & 16+ \\
Sector fixed effects & No & No & Yes & Yes \\
Years & 1992-1998 & 1992-1998 & 1992-1998 & 1992-1998 \\
Observations & 1,197 & 1,197 & 1,197 & 1,197 \\
Sector clusters & 19 & 19 & 19 & 19 \\
R-squared & 0.984 & 0.985 & 0.988 & 0.989 \\
\bottomrule
\end{tabular}
\vspace{6pt}
\begin{minipage}{.96\textwidth}\footnotesize Notes. Standard errors in parentheses. * p<0.10, ** p<0.05, *** p<0.01. Outcome: 100 times net job creation divided by the published DHS average-employment denominator. Coefficients are percentage-point differences from firms aged 16+, which remain in the sample. Ages 0-5 are individual; 6-10 and 11-15 are published bands. Older bands and pre-1977 left-censored firms are pooled by summing flows and denominators. All left-censored firms are at least 16 from 1992. Weights equal the denominator. Macro controls are annual log GDP and nonfarm-business productivity growth times 100. Sector-clustered CR1 covariance; t(18) inference. The entrant rate is approximately 200 percent by definition. Sources: Census BDS 2023 sector-age tables; BEA/BLS series in the source workbook.\end{minipage}
\end{table}

\clearpage
\begin{table}[H]
\centering
\caption{Payroll retention within fixed cohorts: age-five comparisons}\label{tab:A4}
\small
\begin{tabular}{>{\raggedright\arraybackslash}p{0.3600\textwidth}>{\centering\arraybackslash}p{0.1620\textwidth}>{\centering\arraybackslash}p{0.1440\textwidth}>{\centering\arraybackslash}p{0.1170\textwidth}>{\centering\arraybackslash}p{0.1170\textwidth}}
\toprule
Specification & Age-five gap & Std. error & N & Clusters \\
\midrule
Raw & -20.01*** & (1.24) & 1824 & 16 \\
Macro controls & -20.02*** & (1.24) & 1824 & 16 \\
Sector/cohort FE & -20.01*** & (1.25) & 1824 & 16 \\
Both & -18.93*** & (1.17) & 1824 & 16 \\
Equal weights & -17.56*** & (1.50) & 1824 & 16 \\
Sector-clustered & -18.93*** & (5.13) & 1824 & 19 \\
1978-1985 cohorts & -19.24*** & (1.04) & 912 & 8 \\
1986-1993 cohorts & -18.70*** & (1.92) & 912 & 8 \\
Drop recession observations & -18.42*** & (1.67) & 1368 & 16 \\
\bottomrule
\end{tabular}
\vspace{6pt}
\begin{minipage}{.96\textwidth}\footnotesize Notes. Standard errors in parentheses. * p<0.10, ** p<0.05, *** p<0.01. Outcome: 100 times sector-cohort employment divided by that cohort's own age-zero employment. The reported coefficient is the age-five percentage-point difference from entry. Every regression includes indicators for ages 1-5 with age zero omitted. Cohorts enter in 1978-1993 and are followed through age five, ending by 1998. The first four rows add macro controls, sector and birth-cohort effects, and both. Robustness rows include both controls and effects. Baseline weights equal entry employment; covariance is CR1 clustered by birth cohort with t(15) inference. The sector-clustered row uses t(18); subsamples use their own cohort count minus one. These are fixed-cohort retention estimates, not the annual rates in Table~\ref{tab:A3}. Sources: Census BDS; BEA/BLS workbook series.\end{minipage}
\end{table}

\clearpage
\begin{table}[H]
\centering
\caption{Automation-patent exposure and the two employment margins}\label{tab:A5}
\small
\begin{tabular}{>{\raggedright\arraybackslash}p{0.2880\textwidth}>{\centering\arraybackslash}p{0.1530\textwidth}>{\centering\arraybackslash}p{0.1530\textwidth}>{\centering\arraybackslash}p{0.1530\textwidth}>{\centering\arraybackslash}p{0.1530\textwidth}}
\toprule
Outcome & (1) Raw & (2) Size control & (3) Sector FE & (4) Both \\
\midrule
Panel A. Outcome year 1982 &  &  &  &  \\
Pre-1977 origins & \shortstack{-4.83***\\(1.37)} & \shortstack{-4.80***\\(1.29)} & \shortstack{-4.85**\\(1.91)} & \shortstack{-4.95***\\(1.84)} \\
1977-and-later origins & \shortstack{-2.64**\\(1.06)} & \shortstack{-2.72***\\(1.01)} & \shortstack{0.66\\(1.48)} & \shortstack{0.82\\(1.28)} \\
All origins & \shortstack{-7.47***\\(1.34)} & \shortstack{-7.52***\\(1.37)} & \shortstack{-4.19*\\(2.11)} & \shortstack{-4.13*\\(2.13)} \\
Panel B. Outcome year 1998 &  &  &  &  \\
Pre-1977 origins & \shortstack{-17.32***\\(3.42)} & \shortstack{-17.90***\\(2.84)} & \shortstack{-9.65\\(6.20)} & \shortstack{-9.39\\(6.57)} \\
1977-and-later origins & \shortstack{-17.58***\\(5.99)} & \shortstack{-16.92***\\(5.79)} & \shortstack{5.30\\(7.91)} & \shortstack{5.94\\(7.17)} \\
All origins & \shortstack{-34.90***\\(6.47)} & \shortstack{-34.82***\\(6.37)} & \shortstack{-4.35\\(10.04)} & \shortstack{-3.46\\(9.77)} \\
Initial log employment & No & Yes & No & Yes \\
Two-digit sector effects & No & No & Yes & Yes \\
Observations per regression & 84 & 84 & 84 & 84 \\
Sectors represented & 19 & 19 & 19 & 19 \\
\bottomrule
\end{tabular}
\vspace{6pt}
\begin{minipage}{.96\textwidth}\footnotesize Notes. Each cell is a separate weighted least-squares regression. Outcomes are changes from 1978 employment for the stated origin group divided by total 1978 industry employment, multiplied by 100. Exposure is standardized using 1978 employment weights across the 84 industries. Weights are fixed 1978 employment. HC1 standard errors in parentheses; two-sided t tests use residual degrees of freedom. * p<0.10, ** p<0.05, *** p<0.01. Common annual macro variables are constant within each cross-section and absorbed by the intercept. Patent scores combine 1980 composition with 1980-1989 patent flows; STS is an intensity score, not a share. Sources: supplied BDS industry panel, IPUMS employment matrix and Autor et al./DOT scores.\end{minipage}
\end{table}

\clearpage
\begin{table}[H]
\centering
\caption{Routine-cognitive intensity and the two employment margins}\label{tab:A6}
\small
\begin{tabular}{>{\raggedright\arraybackslash}p{0.2880\textwidth}>{\centering\arraybackslash}p{0.1530\textwidth}>{\centering\arraybackslash}p{0.1530\textwidth}>{\centering\arraybackslash}p{0.1530\textwidth}>{\centering\arraybackslash}p{0.1530\textwidth}}
\toprule
Outcome & (1) Raw & (2) Size control & (3) Sector FE & (4) Both \\
\midrule
Panel A. Outcome year 1982 &  &  &  &  \\
Pre-1977 origins & \shortstack{-1.38\\(1.45)} & \shortstack{-1.38\\(1.53)} & \shortstack{1.87\\(2.22)} & \shortstack{1.62\\(2.18)} \\
1977-and-later origins & \shortstack{-2.75***\\(0.98)} & \shortstack{-2.75***\\(0.97)} & \shortstack{-1.03\\(1.47)} & \shortstack{-0.47\\(1.31)} \\
All origins & \shortstack{-4.14***\\(1.42)} & \shortstack{-4.14***\\(1.45)} & \shortstack{0.84\\(1.94)} & \shortstack{1.15\\(1.91)} \\
Panel B. Outcome year 1998 &  &  &  &  \\
Pre-1977 origins & \shortstack{-11.28***\\(3.97)} & \shortstack{-11.28***\\(4.08)} & \shortstack{0.50\\(5.83)} & \shortstack{1.63\\(6.38)} \\
1977-and-later origins & \shortstack{-14.48**\\(5.90)} & \shortstack{-14.47**\\(5.66)} & \shortstack{-2.22\\(8.61)} & \shortstack{0.07\\(7.71)} \\
All origins & \shortstack{-25.76***\\(6.60)} & \shortstack{-25.75***\\(6.69)} & \shortstack{-1.72\\(9.07)} & \shortstack{1.70\\(8.13)} \\
Initial log employment & No & Yes & No & Yes \\
Two-digit sector effects & No & No & Yes & Yes \\
Observations per regression & 84 & 84 & 84 & 84 \\
Sectors represented & 19 & 19 & 19 & 19 \\
\bottomrule
\end{tabular}
\vspace{6pt}
\begin{minipage}{.96\textwidth}\footnotesize Notes. Each cell is a separate weighted least-squares regression. Outcomes are changes from 1978 employment for the stated origin group divided by total 1978 industry employment, multiplied by 100. Exposure is standardized using 1978 employment weights across the 84 industries. Weights are fixed 1978 employment. HC1 standard errors in parentheses; two-sided t tests use residual degrees of freedom. * p<0.10, ** p<0.05, *** p<0.01. Common annual macro variables are constant within each cross-section and absorbed by the intercept. Patent scores combine 1980 composition with 1980-1989 patent flows; STS is an intensity score, not a share. Sources: supplied BDS industry panel, IPUMS employment matrix and Autor et al./DOT scores.\end{minipage}
\end{table}

\clearpage
\begin{table}[H]
\centering
\caption{Annual within-sector exposure gradients, 1979-1998}\label{tab:A7}
\small
\begin{tabular}{>{\raggedright\arraybackslash}p{0.0900\textwidth}>{\centering\arraybackslash}p{0.1350\textwidth}>{\centering\arraybackslash}p{0.1350\textwidth}>{\centering\arraybackslash}p{0.1350\textwidth}>{\centering\arraybackslash}p{0.1350\textwidth}>{\centering\arraybackslash}p{0.1350\textwidth}>{\centering\arraybackslash}p{0.1350\textwidth}}
\toprule
Year & \shortstack{Patent:\\incumbent} & \shortstack{Patent:\\later origins} & \shortstack{Patent:\\total} & \shortstack{Routine:\\incumbent} & \shortstack{Routine:\\later origins} & \shortstack{Routine:\\total} \\
\midrule
1979 & \shortstack{-0.01\\(0.86)} & \shortstack{0.65\\(0.46)} & \shortstack{0.63\\(0.94)} & \shortstack{0.70\\(0.64)} & \shortstack{-0.03\\(0.51)} & \shortstack{0.67\\(0.73)} \\
1980 & \shortstack{-2.23\\(1.41)} & \shortstack{0.48\\(0.77)} & \shortstack{-1.75\\(1.43)} & \shortstack{1.27\\(1.18)} & \shortstack{-0.34\\(0.79)} & \shortstack{0.93\\(0.99)} \\
1981 & \shortstack{-3.00*\\(1.50)} & \shortstack{0.58\\(1.08)} & \shortstack{-2.42*\\(1.44)} & \shortstack{1.11\\(1.52)} & \shortstack{-0.51\\(1.13)} & \shortstack{0.61\\(1.24)} \\
1982 & \shortstack{-4.95***\\(1.84)} & \shortstack{0.82\\(1.28)} & \shortstack{-4.13*\\(2.13)} & \shortstack{1.62\\(2.18)} & \shortstack{-0.47\\(1.31)} & \shortstack{1.15\\(1.91)} \\
1983 & \shortstack{-5.30**\\(2.23)} & \shortstack{1.10\\(1.36)} & \shortstack{-4.20*\\(2.46)} & \shortstack{0.51\\(2.46)} & \shortstack{-0.61\\(1.58)} & \shortstack{-0.10\\(2.23)} \\
1984 & \shortstack{-4.23\\(2.58)} & \shortstack{1.01\\(1.66)} & \shortstack{-3.21\\(2.83)} & \shortstack{2.65\\(2.98)} & \shortstack{-0.31\\(1.83)} & \shortstack{2.34\\(2.70)} \\
1985 & \shortstack{-5.20*\\(2.79)} & \shortstack{2.01\\(1.98)} & \shortstack{-3.19\\(3.28)} & \shortstack{1.92\\(3.11)} & \shortstack{0.15\\(2.22)} & \shortstack{2.08\\(3.04)} \\
1986 & \shortstack{-5.58*\\(3.12)} & \shortstack{2.80\\(2.32)} & \shortstack{-2.78\\(3.76)} & \shortstack{2.07\\(3.43)} & \shortstack{0.75\\(2.55)} & \shortstack{2.81\\(3.32)} \\
1987 & \shortstack{-5.50\\(3.93)} & \shortstack{3.98\\(2.67)} & \shortstack{-1.52\\(4.87)} & \shortstack{1.94\\(3.76)} & \shortstack{2.15\\(2.73)} & \shortstack{4.08\\(3.65)} \\
1988 & \shortstack{-6.24\\(4.25)} & \shortstack{4.92\\(2.99)} & \shortstack{-1.33\\(5.33)} & \shortstack{1.91\\(3.95)} & \shortstack{2.99\\(2.98)} & \shortstack{4.89\\(3.99)} \\
1989 & \shortstack{-5.77\\(4.56)} & \shortstack{5.62\\(3.45)} & \shortstack{-0.15\\(5.76)} & \shortstack{2.12\\(4.11)} & \shortstack{3.40\\(3.47)} & \shortstack{5.52\\(4.31)} \\
1990 & \shortstack{-6.82\\(4.87)} & \shortstack{5.36\\(3.84)} & \shortstack{-1.46\\(6.25)} & \shortstack{2.26\\(4.56)} & \shortstack{3.09\\(3.86)} & \shortstack{5.35\\(4.81)} \\
1991 & \shortstack{-8.65*\\(5.06)} & \shortstack{4.52\\(3.93)} & \shortstack{-4.12\\(6.49)} & \shortstack{1.09\\(5.01)} & \shortstack{2.59\\(4.01)} & \shortstack{3.68\\(5.23)} \\
1992 & \shortstack{-8.18\\(5.29)} & \shortstack{4.25\\(4.23)} & \shortstack{-3.92\\(6.84)} & \shortstack{0.91\\(5.02)} & \shortstack{2.77\\(4.17)} & \shortstack{3.68\\(5.52)} \\
1993 & \shortstack{-8.86\\(5.49)} & \shortstack{4.25\\(4.66)} & \shortstack{-4.61\\(7.20)} & \shortstack{-0.04\\(5.10)} & \shortstack{2.27\\(4.67)} & \shortstack{2.23\\(5.88)} \\
1994 & \shortstack{-8.74\\(5.65)} & \shortstack{4.62\\(5.12)} & \shortstack{-4.13\\(7.59)} & \shortstack{0.31\\(5.42)} & \shortstack{1.94\\(5.21)} & \shortstack{2.25\\(6.38)} \\
1995 & \shortstack{-8.76\\(6.05)} & \shortstack{5.14\\(5.83)} & \shortstack{-3.62\\(8.36)} & \shortstack{0.31\\(5.70)} & \shortstack{1.74\\(5.99)} & \shortstack{2.05\\(7.02)} \\
1996 & \shortstack{-9.01\\(6.32)} & \shortstack{5.79\\(6.23)} & \shortstack{-3.22\\(8.90)} & \shortstack{0.15\\(5.92)} & \shortstack{1.51\\(6.62)} & \shortstack{1.67\\(7.60)} \\
1997 & \shortstack{-9.53\\(6.41)} & \shortstack{5.79\\(6.82)} & \shortstack{-3.74\\(9.58)} & \shortstack{-0.11\\(6.00)} & \shortstack{0.73\\(7.28)} & \shortstack{0.63\\(8.21)} \\
1998 & \shortstack{-9.39\\(6.57)} & \shortstack{5.94\\(7.17)} & \shortstack{-3.46\\(9.77)} & \shortstack{1.63\\(6.38)} & \shortstack{0.07\\(7.71)} & \shortstack{1.70\\(8.13)} \\
\bottomrule
\end{tabular}
\vspace{6pt}
\begin{minipage}{.96\textwidth}\footnotesize Notes. Outcome units, exposure standardization, and weights match Tables A5-A6. All columns control for initial log employment and two-digit sector effects. HC1 standard errors in parentheses; pointwise two-sided t(63) inference. * p<0.10, ** p<0.05, *** p<0.01. Each regression has 84 observations. At 1978 every outcome is zero by construction. The three slopes for each exposure sum exactly: incumbent + later origins = total.\end{minipage}
\end{table}

\clearpage
\begin{table}[H]
\centering
\caption{Industry-exposure robustness at the 1998 endpoint}\label{tab:A8}
\small
\begin{tabular}{>{\raggedright\arraybackslash}p{0.3420\textwidth}>{\centering\arraybackslash}p{0.1530\textwidth}>{\centering\arraybackslash}p{0.1530\textwidth}>{\centering\arraybackslash}p{0.1530\textwidth}>{\centering\arraybackslash}p{0.0990\textwidth}}
\toprule
Specification & Incumbent & Later origins & Total & N \\
\midrule
HC1 baseline & \shortstack{-9.39\\(6.57)} & \shortstack{5.94\\(7.17)} & \shortstack{-3.46\\(9.77)} & 84 \\
Sector-clustered CR1 & \shortstack{-9.39\\(6.55)} & \shortstack{5.94\\(5.11)} & \shortstack{-3.46\\(6.72)} & 84 \\
Equal industry weights & \shortstack{26.27\\(35.12)} & \shortstack{2.52\\(8.22)} & \shortstack{28.79\\(39.78)} & 84 \\
Exclude manufacturing & \shortstack{-9.61\\(7.17)} & \shortstack{7.51\\(7.64)} & \shortstack{-2.10\\(10.53)} & 63 \\
Add augmentation exposure & \shortstack{-0.87\\(9.44)} & \shortstack{21.44\\(19.18)} & \shortstack{20.58\\(19.42)} & 84 \\
Routine-cognitive intensity & \shortstack{1.63\\(6.38)} & \shortstack{0.07\\(7.71)} & \shortstack{1.70\\(8.13)} & 84 \\
Overall routine intensity & \shortstack{0.39\\(6.22)} & \shortstack{2.72\\(8.14)} & \shortstack{3.11\\(9.26)} & 84 \\
Routine-manual intensity & \shortstack{-1.09\\(5.17)} & \shortstack{5.51\\(9.74)} & \shortstack{4.41\\(11.45)} & 84 \\
HC3; omit singleton sectors & \shortstack{-9.39\\(9.32)} & \shortstack{5.94\\(10.20)} & \shortstack{-3.46\\(13.74)} & 80 \\
Change from 1985 to 1998 & \shortstack{-4.19\\(4.54)} & \shortstack{3.93\\(5.52)} & \shortstack{-0.27\\(7.50)} & 84 \\
\bottomrule
\end{tabular}
\vspace{6pt}
\begin{minipage}{.96\textwidth}\footnotesize Notes. All regressions contain initial log employment and two-digit sector effects. The main exposure is automation patents unless the row names an alternative. All exposures retain the full-sample weighted normalization. Weights are 1978 employment except in the equal-weight row. Parentheses contain the named covariance estimate (otherwise HC1); cluster tests use t(G-1), other tests residual t degrees of freedom. HC3 omits sectors with a single industry to avoid unit leverage; the point estimates are unchanged. The last row subtracts 1985 outcomes, with the 1978 denominator, weights and exposure held fixed. * p<0.10, ** p<0.05, *** p<0.01.\end{minipage}
\end{table}

\clearpage
\begin{table}[H]
\centering
\caption{Calibration and economic interpretation}\label{tab:A9}
\small
\begin{tabular}{>{\raggedright\arraybackslash}p{0.2520\textwidth}>{\centering\arraybackslash}p{0.1170\textwidth}>{\centering\arraybackslash}p{0.1890\textwidth}>{\centering\arraybackslash}p{0.3420\textwidth}}
\toprule
Parameter & Benchmark & Sensitivity & Interpretation \\
\midrule
Product substitution, epsilon & 4 & 4; 6 & CES benchmark; markup 33\%; 20\% \\
Hired-hours elasticity, phi & 0.50 & 0.25; 0.50 & 0.50: intensive-margin Frisch benchmark \\
Founder-effort elasticity, \(\phi_F\) & 0.50 & 0.25; 0.50 & Scenario; not an estimated founder elasticity \\
Consumption curvature, sigma & 0.6 & 0; 0.5; 0.75; 1 & Partial income effects; sigma=1 stronger sensitivity \\
Compute supply, eta & 1 & 0.25; 1; 4 & Quantity response to its price \\
Founder development share, theta & 0.80 & 0.50; 0.80; 1 & Cost share in development, not worker share \\
Discount rate, rho & 0.04/year & Fixed & Annual impatience \\
Product exit; abandonment & 0.05/year each & Fixed & Separate exit hazards \\
Development curvature, xi & 0.50 & Fixed & Diminishing returns to development effort \\
Development scale, \(\nu_0\) & 0.50 & 0.25; 0.50; 1 & Initial completion/exit interval: 2.7-7.7 years \\
Human review, v & 0.20 & Fixed & Per accepted AI-assisted task \\
Task compute requirement & Uniform [0,2] & Fixed & Task heterogeneity in normalized units \\
Maintenance, \(f_m\) & 0.05 & Fixed & Hired hours per active product \\
Initial M; AI share; effort d & 1; 0.20; 0.10 & Anchors & Effective units; not literal startup headcount \\
Initial founder opportunity cost & \(w_F\) = w & Anchor & Separate prices after the shock \\
Compute efficiency, gamma\_1/gamma\_0 & 1.50 & 1.25; 1.50; 2 & Efficiency shock; also isolated channel run \\
Linear initiation cost, b\_1/b\_0 & 0.75 & 0.50; 0.75; 0.90 & Only linear term; b\_0=0.60 \\
Initiation curvature, zeta & 7.7663 & Recovered & Project-value/entry condition \\
Compute scale, kappa & 0.0211 & Recovered & Initial compute quantity and price \\
Hired disutility scale, \(\chi_H\) & 1.4997 & Recovered & Initial hired-hours supply \\
Founder disutility scale, \(\chi_F\) & 136.6746 & Recovered & Initial founder-effort supply \\
Incumbent scale speed, \(\lambda\) (extension) & 0.20/year & 0; .10; .35; .70; infinite & Fixed-target half-life 3.47 years \\
Efficiency diffusion (extension) & Immediate & Normalized logistic & Location 3.5 years; scale 1.2 years \\
\bottomrule
\end{tabular}
\vspace{6pt}
\begin{minipage}{.96\textwidth}\footnotesize Notes. Parameters are annualized where indicated. Product substitution is guided by Bilbiie, Ghironi, and Melitz (2012); the 0.50 hired-hours benchmark is guided by Chetty et al. (2013). Founder shares, supply responsiveness, effort technology, and shocks define scenarios, not estimated confidence intervals. Recovered scales are recalibrated in each comparison to preserve the stated initial anchors. Founder time is valued at its opportunity cost; it is not recorded as a salary. The extension changes the transition, not the initial allocation. A normalized logistic efficiency path is distinct from endogenous task adoption.\end{minipage}
\end{table}

\clearpage
\begin{table}[H]
\centering
\caption{Year-15 equilibrium outcomes across scenarios}\label{tab:A10}
\small
\begin{tabular}{>{\raggedright\arraybackslash}p{0.3060\textwidth}>{\centering\arraybackslash}p{0.1188\textwidth}>{\centering\arraybackslash}p{0.1188\textwidth}>{\centering\arraybackslash}p{0.1188\textwidth}>{\centering\arraybackslash}p{0.1188\textwidth}>{\centering\arraybackslash}p{0.1188\textwidth}}
\toprule
Scenario & Hired hours & Founder effort & Real wage & Consumption & Products \\
\midrule
Benchmark (sigma=0.6) & 0.27 & 0.69 & 4.41 & 6.51 & 5.03 \\
No income effects (sigma=0) & 2.31 & 3.29 & 4.67 & 8.96 & 6.02 \\
Stronger income effects (sigma=1) & -0.57 & -0.36 & 4.30 & 5.51 & 4.63 \\
Tight compute (eta=.25) & 0.15 & 0.22 & 3.69 & 5.70 & 4.82 \\
Elastic compute (eta=4) & 0.49 & 1.60 & 5.78 & 8.03 & 5.44 \\
Founder-only development & 0.24 & 0.85 & 4.35 & 6.50 & 4.86 \\
Mixed development (theta=.5) & 0.31 & 0.38 & 4.50 & 6.51 & 5.28 \\
Lower hours elasticities & 0.14 & 0.38 & 4.34 & 6.32 & 4.81 \\
Product substitution=6 & 0.16 & 0.57 & 3.69 & 5.67 & 5.00 \\
Slower completion & 0.18 & 1.07 & 4.04 & 6.17 & 3.97 \\
Faster completion & 0.35 & 0.22 & 4.77 & 6.83 & 6.09 \\
Smaller AI shock & 0.10 & 0.47 & 2.05 & 3.09 & 2.00 \\
Larger AI shock & 0.54 & 0.82 & 8.90 & 13.21 & 10.77 \\
Compute-efficiency channel only & -0.03 & 1.30 & 2.86 & 4.92 & 0.59 \\
Initiation-cost channel only & 0.30 & -0.62 & 1.49 & 1.49 & 4.41 \\
Partial income effect: 0.5 & 0.52 & 1.01 & 4.44 & 6.82 & 5.16 \\
Partial income effect: 0.75 & -0.08 & 0.25 & 4.37 & 6.09 & 4.87 \\
\bottomrule
\end{tabular}
\vspace{6pt}
\begin{minipage}{.96\textwidth}\footnotesize Notes. Each cell is a percentage change from that scenario's initial steady state. Founder effort includes initiation and own development time; hired hours include operating production, review, maintenance, and hired development. Products are active varieties, not employees or legal firm counts. The benchmark jointly raises compute efficiency 50\% and lowers the linear initiation requirement 25\%. Smaller/larger shocks use 25\%/100\% efficiency improvements and 10\%/50\% initiation savings. All other comparisons vary the named parameter around the sigma=0.6 benchmark. These are one-at-a-time scenario results, not confidence intervals. Partial-income-effect cases vary only the named consumption curvature and the corresponding calibrated scale parameters.\end{minipage}
\end{table}

\clearpage
\begin{table}[H]
\centering
\caption{Incumbent-scale adjustment: impact and recovery}\label{tab:A11}
\small
\begin{tabular}{>{\raggedright\arraybackslash}p{0.2790\textwidth}>{\centering\arraybackslash}p{0.1260\textwidth}>{\centering\arraybackslash}p{0.1260\textwidth}>{\centering\arraybackslash}p{0.1260\textwidth}>{\centering\arraybackslash}p{0.1170\textwidth}>{\centering\arraybackslash}p{0.1260\textwidth}}
\toprule
Scenario & Hours: impact & Wage: impact & Hours: minimum & Trough year & Recovery year \\
\midrule
Instant scaling (reference) & -0.07 & 2.65 & -0.07 & 0.0 & 4.1 \\
\(\lambda\) = 0.7 & -4.61 & -9.29 & -4.61 & 0.0 & 2.8 \\
\(\lambda\) = 0.35 & -4.60 & -9.26 & -4.60 & 0.0 & 2.5 \\
\(\lambda\) = 0.2 & -4.58 & -9.23 & -4.58 & 0.0 & 3.0 \\
\(\lambda\) = 0.1 & -4.55 & -9.18 & -4.55 & 0.0 & 4.4 \\
\(\lambda\) = 0 & -4.46 & -9.02 & -4.46 & 0.0 & 11.1 \\
\(\lambda\) = 0.2; efficiency diffusing over seven years & 0.01 & 0.02 & -0.23 & 3.8 & 6.8 \\
\(\lambda\) = 0; efficiency diffusing over seven years & 0.02 & 0.04 & -0.59 & 4.7 & 11.1 \\
Instant scaling; efficiency diffusing over seven years & -0.00 & -0.00 & -0.00 & 0.0 & 0.1 \\
\(\lambda\) = 0.2; sigma = 0 & -4.55 & -8.90 & -4.55 & 0.0 & 0.9 \\
\(\lambda\) = 0.2; sigma = 1 & -4.59 & -9.42 & -4.59 & 0.0 & None by 25 \\
\bottomrule
\end{tabular}
\vspace{6pt}
\begin{minipage}{.96\textwidth}\footnotesize Notes. Model simulations; percentage changes from the scenario initial equilibrium. Recovery is the first upward zero crossing after a negative excursion, interpolated on a 0.1-year grid. The infinite-speed case is solved using the exact symmetric model. All scenarios use sigma=0.6 unless noted. The diffusing-efficiency case uses the normalized logistic path; founder effort remains separate from hired hours.\end{minipage}
\end{table}

\clearpage
\begin{table}[H]
\centering
\caption{Incumbent-scale adjustment: year-15 outcomes}\label{tab:A12}
\small
\begin{tabular}{>{\raggedright\arraybackslash}p{0.2430\textwidth}>{\centering\arraybackslash}p{0.0900\textwidth}>{\centering\arraybackslash}p{0.0900\textwidth}>{\centering\arraybackslash}p{0.0990\textwidth}>{\centering\arraybackslash}p{0.0900\textwidth}>{\centering\arraybackslash}p{0.0990\textwidth}>{\centering\arraybackslash}p{0.0900\textwidth}>{\centering\arraybackslash}p{0.0990\textwidth}}
\toprule
Scenario & Hired hours & Existing & Later + develop. & Real wage & Consumption & Products & Founder effort \\
\midrule
Instant scaling (reference) & 0.27 & -2.16 & 2.42 & 4.41 & 6.51 & 5.03 & 0.69 \\
\(\lambda\) = 0.7 & 0.28 & -2.04 & 2.32 & 4.45 & 6.52 & 5.01 & 0.66 \\
\(\lambda\) = 0.35 & 0.30 & -1.92 & 2.22 & 4.49 & 6.53 & 4.98 & 0.64 \\
\(\lambda\) = 0.2 & 0.32 & -1.75 & 2.08 & 4.55 & 6.55 & 4.94 & 0.60 \\
\(\lambda\) = 0.1 & 0.36 & -1.46 & 1.83 & 4.66 & 6.58 & 4.87 & 0.52 \\
\(\lambda\) = 0 & 0.22 & -2.53 & 2.75 & 4.30 & 6.50 & 5.17 & 0.56 \\
\(\lambda\) = 0.2; efficiency diffusing over seven years & 0.32 & -1.80 & 2.12 & 4.55 & 6.56 & 4.97 & 0.60 \\
\(\lambda\) = 0; efficiency diffusing over seven years & 0.22 & -2.53 & 2.75 & 4.30 & 6.50 & 5.17 & 0.56 \\
Instant scaling; efficiency diffusing over seven years & 0.27 & -2.16 & 2.42 & 4.41 & 6.51 & 5.03 & 0.69 \\
\(\lambda\) = 0.2; sigma = 0 & 2.38 & -1.24 & 3.62 & 4.82 & 9.02 & 5.92 & 3.20 \\
\(\lambda\) = 0.2; sigma = 1 & -0.52 & -1.97 & 1.45 & 4.44 & 5.54 & 4.53 & -0.44 \\
\bottomrule
\end{tabular}
\vspace{6pt}
\begin{minipage}{.96\textwidth}\footnotesize Notes. Existing and later-product contributions use initial hired hours as a common denominator and sum to the hired-hours change. Other outcomes are percentage changes from the initial equilibrium. Products are varieties, not legal firms or worker counts. \(\lambda\) changes transition dynamics but not the limiting steady state as inherited varieties exit.\end{minipage}
\end{table}

\clearpage
\begin{table}[H]
\centering
\caption{Adjustment speed, diffusion, and the depth of the dip}\label{tab:A13}
\small
\begin{tabular}{>{\raggedright\arraybackslash}p{0.3150\textwidth}>{\centering\arraybackslash}p{0.1170\textwidth}>{\centering\arraybackslash}p{0.1170\textwidth}>{\centering\arraybackslash}p{0.1170\textwidth}>{\centering\arraybackslash}p{0.1170\textwidth}>{\centering\arraybackslash}p{0.1170\textwidth}}
\toprule
Incumbent scale speed & Sudden & 2 years & 5 years & 7 years & 10 years \\
\midrule
Panel A. Trough in aggregate hired hours (percent below baseline) &  &  &  &  &  \\
\(\lambda\) = 0 & 4.46 & 1.91 & 0.91 & 0.59 & 0.26 \\
\(\lambda\) = 0.1 & 4.55 & 1.40 & 0.57 & 0.34 & 0.12 \\
\(\lambda\) = 0.2 (benchmark) & 4.58 & 1.09 & 0.40 & 0.23 & 0.07 \\
\(\lambda\) = 0.35 & 4.60 & 0.80 & 0.28 & 0.15 & 0.03 \\
\(\lambda\) = 0.7 & 4.61 & 0.50 & 0.15 & 0.07 & 0.02 \\
\(\lambda\) infinite (instant re-scaling) & 0.07 & 0.05 & 0.00 & 0.00 & 0.00 \\
Panel B. Year hired hours return to baseline &  &  &  &  &  \\
\(\lambda\) = 0 & 11.1 & 11.1 & 11.1 & 11.2 & 11.1 \\
\(\lambda\) = 0.1 & 4.4 & 5.6 & 7.1 & 7.9 & 8.4 \\
\(\lambda\) = 0.2 (benchmark) & 3.1 & 4.3 & 6.1 & 6.9 & 7.3 \\
\(\lambda\) = 0.35 & 2.5 & 3.6 & 5.5 & 6.2 & 6.2 \\
\(\lambda\) = 0.7 & 2.9 & 3.3 & 4.9 & 5.4 & 3.6 \\
\(\lambda\) infinite (instant re-scaling) & 4.1 & 4.1 & 3.3 & 0.6 & 0.6 \\
\bottomrule
\end{tabular}
\vspace{6pt}
\begin{minipage}{.96\textwidth}\footnotesize Notes. Each cell is a simulation of the extension in Section 3.6 with sigma=0.6, the benchmark 50 percent compute-efficiency improvement and 25 percent initiation-cost reduction, and the stated incumbent scale-adjustment rate. Columns give the arrival of the efficiency gain: all at once, or along the logistic path of \hyperref[app:D]{Appendix D} scaled so that the gain moves from 5 to 95 percent of its final value over the stated number of years; the seven-year column is the case in Figure~\ref{fig:11}. Panel A reports the largest decline in aggregate hired hours relative to the no-shock path; Panel B the first year at which hired hours return to that path after the trough. The initiation-cost improvement arrives at once in every case. Troughs below about 0.5 percent are within the resolution of the payroll evidence in Section 5; troughs above 2 percent would have appeared as a decline.\end{minipage}
\end{table}

\clearpage
\begin{table}[H]
\centering
\caption{LLM exposure and establishment job flows after ChatGPT}\label{tab:A14}
\small
\begin{tabular}{>{\raggedright\arraybackslash}p{0.3150\textwidth}>{\centering\arraybackslash}p{0.1463\textwidth}>{\centering\arraybackslash}p{0.1463\textwidth}>{\centering\arraybackslash}p{0.1463\textwidth}>{\centering\arraybackslash}p{0.1463\textwidth}}
\toprule
Outcome & (1) Raw & (2) Size control & (3) Sector FE & (4) Both \\
\midrule
Panel A. Outcome quarter 2023Q4 &  &  &  &  \\
Opening establishments & \shortstack{-0.29\\(0.67)} & \shortstack{-0.26\\(0.46)} & \shortstack{0.04\\(0.57)} & \shortstack{0.06\\(0.56)} \\
Continuing establishments, net & \shortstack{-0.29\\(0.69)} & \shortstack{-0.30\\(0.57)} & \shortstack{-0.39\\(0.80)} & \shortstack{-0.45\\(0.76)} \\
All establishments & \shortstack{-0.57**\\(0.27)} & \shortstack{-0.56**\\(0.27)} & \shortstack{-0.35\\(0.42)} & \shortstack{-0.39\\(0.42)} \\
Panel B. Outcome quarter 2026Q1 &  &  &  &  \\
Opening establishments & \shortstack{-0.98\\(2.18)} & \shortstack{-0.89\\(1.45)} & \shortstack{0.18\\(1.94)} & \shortstack{0.24\\(1.91)} \\
Continuing establishments, net & \shortstack{0.19\\(2.42)} & \shortstack{0.13\\(1.95)} & \shortstack{-0.94\\(1.94)} & \shortstack{-1.09\\(1.85)} \\
All establishments & \shortstack{-0.79\\(0.62)} & \shortstack{-0.76\\(0.77)} & \shortstack{-0.76\\(0.92)} & \shortstack{-0.85\\(0.98)} \\
Panel C. Marchesi-Tang top versus bottom 20, outcome quarter 2026Q1 &  &  &  &  \\
Opening establishments & \shortstack{6.30\\(5.19)} & \shortstack{5.10\\(5.75)} & -- & -- \\
Continuing establishments, net & \shortstack{-15.81**\\(6.27)} & \shortstack{-15.72**\\(6.73)} & -- & -- \\
All establishments & \shortstack{-9.50***\\(2.24)} & \shortstack{-10.61***\\(2.03)} & -- & -- \\
Initial log employment & No & Yes & No & Yes \\
Two-digit sector effects & No & No & Yes & Yes \\
Observations per regression (Panels A-B; Panel C) & 72; 27 & 72; 27 & 72 & 72 \\
Sectors represented & 19 & 19 & 19 & 19 \\
\bottomrule
\end{tabular}
\vspace{6pt}
\begin{minipage}{.96\textwidth}\footnotesize Notes. Each cell is a separate weighted least-squares regression across three-digit NAICS industries. Outcomes are cumulative job flows from 2022Q4 to the stated quarter as a percent of 2019 industry employment: job gains at opening establishments; expansions minus contractions minus closings at continuing establishments; and their sum. Panels A and B use LLM exposure measured as the employment-weighted mean of Eloundou et al. (2024) GPT-4 beta exposure over the industry's May 2022 occupational composition (BLS OEWS), per weighted standard deviation; weights are 2019 BED employment. Panel C uses an indicator for the top 20 versus bottom 20 of the 81 industries in the published table of Marchesi and Tang (2026), the only scores they report; 27 of the 40 are BED industries, and sector effects cannot be included. HC1 standard errors in parentheses; * p<0.10, ** p<0.05, *** p<0.01. Sources: BLS Business Employment Dynamics (October 28, 2026 release); OEWS May 2022; Eloundou et al. (2024) replication data; Marchesi and Tang (2026).\end{minipage}
\end{table}

\clearpage
\begin{table}[H]
\centering
\caption{Measurement lags and the recorded path of hired hours}\label{tab:A15}
\small
\begin{tabular}{>{\raggedright\arraybackslash}p{0.3150\textwidth}>{\centering\arraybackslash}p{0.1463\textwidth}>{\centering\arraybackslash}p{0.1463\textwidth}>{\centering\arraybackslash}p{0.1463\textwidth}>{\centering\arraybackslash}p{0.1463\textwidth}}
\toprule
Recording of new-activity hours & Trough (percent) & Trough year & Return to baseline (year) & Year 15 (percent) \\
\midrule
Panel A. \(\lambda\) = 0.2, sudden efficiency gain &  &  &  &  \\
No lag (true path) & 4.58 & 0.0 & 3.1 & 0.32 \\
Mean lag 1 year & 4.75 & 0.0 & 7.2 & 0.26 \\
Mean lag 2 years & 4.75 & 0.0 & 11.5 & 0.16 \\
Mean lag 3 years & 4.75 & 0.0 & 14.2 & 0.05 \\
Panel B. \(\lambda\) = 0.2, efficiency diffusing over seven years &  &  &  &  \\
No lag (true path) & 0.23 & 3.8 & 6.9 & 0.32 \\
Mean lag 1 year & 0.60 & 3.8 & 7.7 & 0.27 \\
Mean lag 2 years & 0.84 & 4.1 & 9.3 & 0.21 \\
Mean lag 3 years & 0.99 & 4.3 & 11.7 & 0.13 \\
Panel C. \(\lambda\) = 0, efficiency diffusing over seven years &  &  &  &  \\
No lag (true path) & 0.59 & 4.7 & 11.2 & 0.22 \\
Mean lag 1 year & 1.24 & 4.4 & 10.7 & 0.30 \\
Mean lag 2 years & 1.69 & 4.7 & 11.3 & 0.35 \\
Mean lag 3 years & 1.99 & 4.9 & 12.2 & 0.33 \\
\bottomrule
\end{tabular}
\vspace{6pt}
\begin{minipage}{.96\textwidth}\footnotesize Notes. Simulated paths from Section 4.3 (Tables A11-A12). Hours at products operating before the shock are recorded without delay. Hours in later products and hired development enter the recorded series through an exponential delay with the stated mean, representing the time for founder time, contractor work, and platform work to appear as payroll. The true paths are unchanged; the table reports the series a statistical system with that delay would show. Trough and year-15 values are percent of initial hired hours.\end{minipage}
\end{table}

\clearpage
\end{document}